\documentclass[10pt,journal,compsoc,twoside]{IEEEtran}

\RequirePackage{xcolor}
\PassOptionsToPackage{binary-units, detect-all, per-mode=symbol, range-units=single, range-phrase=~to~}{siunitx}
\RequirePackage{siunitx}
\PassOptionsToPackage{acronyms,nohypertypes={acronym},nomain}{glossaries}
\RequirePackage{glossaries}
\RequirePackage[pscoord]{eso-pic}
\RequirePackage{multirow}
\RequirePackage{makecell}
\RequirePackage{subfig}
\RequirePackage{threeparttable}
\RequirePackage{booktabs}
\RequirePackage{colortbl}
\RequirePackage[verbose]{placeins}
\RequirePackage{fancyhdr}
\RequirePackage[hidelinks]{hyperref}
\RequirePackage{MnSymbol}
\RequirePackage{wasysym}
\RequirePackage{xspace}
\RequirePackage{lipsum}
\RequirePackage{soul}
\RequirePackage{calc}

\ifCLASSOPTIONcompsoc
  \usepackage[nocompress]{cite}
\else
  \usepackage{cite}
\fi

\usepackage[pdftex]{graphicx}
\RequirePackage{orcidlink}
\RequirePackage[noabbrev,capitalise]{cleveref}

\usepackage{algorithmic}

\usepackage{array}

\usepackage{url}

\newcommand{\glsf}[1]{\glsreset{#1}\gls{#1}}

\newcommand{\Glsplf}[1]{\glsreset{#1}\Glspl{#1}}
\newcommand{\glsu}[1]{\glsunset{#1}\gls{#1}}

\newcommand{\etal}{\emph{et al.}}
\newcommand{\x}{$\times$}

\DeclareSIUnit{\x}{\!\ensuremath{\times}}
\DeclareSIUnit\bit{b}
\DeclareSIUnit\gateeq{GE}
\newlength\myheight
\newlength\mydepth
\settototalheight\myheight{Xygp}
\newcommand*\inlinegraphics[1]{%
    \settototalheight\myheight{Xygp}%
    \settodepth\mydepth{Xygp}%
    \hspace{-2pt}\raisebox{-\mydepth}{\includegraphics[height=\myheight]{#1}}\hspace{-8pt} %
}
\newcommand*\inlinegraphicscap[1]{%
    \protect\inlinegraphics{#1}\protect
}

\definecolor{ieee-bright-dblue-100}{rgb}{0.0, 0.3828, 0.6055}
\definecolor{ieee-bright-dblue-80}{rgb}{0.0, 0.4883, 0.6797}
\definecolor{ieee-bright-dblue-60}{rgb}{0.3633, 0.6094, 0.7617}
\definecolor{ieee-bright-dblue-40}{rgb}{0.5898, 0.7383, 0.8398}
\definecolor{ieee-bright-dblue-20}{rgb}{0.8906, 0.8984, 0.9219}
\definecolor{ieee-bright-red-100}{rgb}{0.7266, 0.0469, 0.1836}
\definecolor{ieee-bright-red-80}{rgb}{0.832, 0.3164, 0.3281}
\definecolor{ieee-bright-red-60}{rgb}{0.8906, 0.4922, 0.4805}
\definecolor{ieee-bright-red-40}{rgb}{0.9336, 0.6562, 0.6406}
\definecolor{ieee-bright-red-20}{rgb}{0.9688, 0.8203, 0.8125}
\definecolor{ieee-bright-orange-100}{rgb}{0.9961, 0.6367, 0.0}
\definecolor{ieee-bright-orange-80}{rgb}{0.9844, 0.6953, 0.3125}
\definecolor{ieee-bright-orange-60}{rgb}{0.9883, 0.7695, 0.4844}
\definecolor{ieee-bright-orange-40}{rgb}{0.9922, 0.8359, 0.6562}
\definecolor{ieee-bright-orange-20}{rgb}{0.9961, 0.9219, 0.8164}
\definecolor{ieee-bright-yellow-100}{rgb}{0.9961, 0.8164, 0.0}
\definecolor{ieee-bright-yellow-80}{rgb}{0.9961, 0.8477, 0.2148}
\definecolor{ieee-bright-yellow-60}{rgb}{0.9961, 0.875, 0.4492}
\definecolor{ieee-bright-yellow-40}{rgb}{0.9961, 0.9062, 0.6328}
\definecolor{ieee-bright-yellow-20}{rgb}{0.9961, 0.9531, 0.8125}
\definecolor{ieee-bright-lgreen-100}{rgb}{0.4688, 0.7422, 0.125}
\definecolor{ieee-bright-lgreen-80}{rgb}{0.5742, 0.7852, 0.332}
\definecolor{ieee-bright-lgreen-60}{rgb}{0.6875, 0.8398, 0.5039}
\definecolor{ieee-bright-lgreen-40}{rgb}{0.793, 0.8906, 0.6641}
\definecolor{ieee-bright-lgreen-20}{rgb}{0.8945, 0.9414, 0.8281}
\definecolor{ieee-bright-dgreen-100}{rgb}{0.0, 0.5156, 0.2383}
\definecolor{ieee-bright-dgreen-80}{rgb}{0.1641, 0.6055, 0.3867}
\definecolor{ieee-bright-dgreen-60}{rgb}{0.3906, 0.6953, 0.5234}
\definecolor{ieee-bright-dgreen-40}{rgb}{0.6094, 0.8008, 0.6719}
\definecolor{ieee-bright-dgreen-20}{rgb}{0.8047, 0.8945, 0.8359}
\definecolor{ieee-bright-purple-100}{rgb}{0.5938, 0.1133, 0.5898}
\definecolor{ieee-bright-purple-80}{rgb}{0.6992, 0.3281, 0.668}
\definecolor{ieee-bright-purple-60}{rgb}{0.7812, 0.4961, 0.7461}
\definecolor{ieee-bright-purple-40}{rgb}{0.8555, 0.6602, 0.8281}
\definecolor{ieee-bright-purple-20}{rgb}{0.9219, 0.8281, 0.9023}
\definecolor{ieee-bright-lblue-100}{rgb}{0.0, 0.6094, 0.6484}
\definecolor{ieee-bright-lblue-80}{rgb}{0.0, 0.6797, 0.7188}
\definecolor{ieee-bright-lblue-60}{rgb}{0.2109, 0.75, 0.7812}
\definecolor{ieee-bright-lblue-40}{rgb}{0.5469, 0.8242, 0.8438}
\definecolor{ieee-bright-lblue-20}{rgb}{0.7695, 0.918, 0.9219}
\definecolor{ieee-bright-cyan-100}{rgb}{0.0, 0.707, 0.8828}
\definecolor{ieee-bright-cyan-80}{rgb}{0.0, 0.7227, 0.9453}
\definecolor{ieee-bright-cyan-60}{rgb}{0.2656, 0.7812, 0.957}
\definecolor{ieee-bright-cyan-40}{rgb}{0.5547, 0.8438, 0.9688}
\definecolor{ieee-bright-cyan-20}{rgb}{0.7773, 0.9141, 0.9805}
\definecolor{ieee-bright-white-100}{rgb}{0.9961, 0.9961, 0.9961}
\definecolor{ieee-bright-white-80}{rgb}{0.9961, 0.9961, 0.9961}
\definecolor{ieee-bright-white-60}{rgb}{0.9961, 0.9961, 0.9961}
\definecolor{ieee-bright-white-40}{rgb}{0.9961, 0.9961, 0.9961}
\definecolor{ieee-bright-white-20}{rgb}{0.9961, 0.9961, 0.9961}
\definecolor{ieee-dark-red-100}{rgb}{0.5234, 0.1211, 0.2539}
\definecolor{ieee-dark-red-80}{rgb}{0.6445, 0.2812, 0.3828}
\definecolor{ieee-dark-red-60}{rgb}{0.7422, 0.4727, 0.5234}
\definecolor{ieee-dark-red-40}{rgb}{0.832, 0.6445, 0.6758}
\definecolor{ieee-dark-red-20}{rgb}{0.918, 0.8203, 0.832}
\definecolor{ieee-dark-orange-100}{rgb}{0.9062, 0.4648, 0.1328}
\definecolor{ieee-dark-orange-80}{rgb}{0.9648, 0.5664, 0.3164}
\definecolor{ieee-dark-orange-60}{rgb}{0.9766, 0.6758, 0.4805}
\definecolor{ieee-dark-orange-40}{rgb}{0.9844, 0.7773, 0.6523}
\definecolor{ieee-dark-orange-20}{rgb}{0.9922, 0.8789, 0.8125}
\definecolor{ieee-dark-yellow-100}{rgb}{0.9961, 0.7773, 0.1719}
\definecolor{ieee-dark-yellow-80}{rgb}{0.9961, 0.8086, 0.375}
\definecolor{ieee-dark-yellow-60}{rgb}{0.9961, 0.875, 0.4492}
\definecolor{ieee-dark-yellow-40}{rgb}{0.9961, 0.8984, 0.6875}
\definecolor{ieee-dark-yellow-20}{rgb}{0.9961, 0.9453, 0.8438}
\definecolor{ieee-dark-lgreen-100}{rgb}{0.3945, 0.5508, 0.0938}
\definecolor{ieee-dark-lgreen-80}{rgb}{0.5078, 0.6289, 0.293}
\definecolor{ieee-dark-lgreen-60}{rgb}{0.6367, 0.7188, 0.4688}
\definecolor{ieee-dark-lgreen-40}{rgb}{0.7539, 0.8047, 0.6367}
\definecolor{ieee-dark-lgreen-20}{rgb}{0.875, 0.9023, 0.8125}
\definecolor{ieee-dark-dgreen-100}{rgb}{0.0, 0.3867, 0.2539}
\definecolor{ieee-dark-dgreen-80}{rgb}{0.1836, 0.5, 0.3906}
\definecolor{ieee-dark-dgreen-60}{rgb}{0.3984, 0.6172, 0.5273}
\definecolor{ieee-dark-dgreen-40}{rgb}{0.5938, 0.7422, 0.6758}
\definecolor{ieee-dark-dgreen-20}{rgb}{0.793, 0.8711, 0.8359}
\definecolor{ieee-dark-purple-100}{rgb}{0.4648, 0.1445, 0.5117}
\definecolor{ieee-dark-purple-80}{rgb}{0.5898, 0.3242, 0.6016}
\definecolor{ieee-dark-purple-60}{rgb}{0.6914, 0.4883, 0.6953}
\definecolor{ieee-dark-purple-40}{rgb}{0.7969, 0.6523, 0.793}
\definecolor{ieee-dark-purple-20}{rgb}{0.8945, 0.8203, 0.8945}
\definecolor{ieee-dark-cyan-100}{rgb}{0.0, 0.4492, 0.4648}
\definecolor{ieee-dark-cyan-80}{rgb}{0.0, 0.5469, 0.5664}
\definecolor{ieee-dark-cyan-60}{rgb}{0.3047, 0.6602, 0.668}
\definecolor{ieee-dark-cyan-40}{rgb}{0.5586, 0.7695, 0.7734}
\definecolor{ieee-dark-cyan-20}{rgb}{0.7734, 0.8789, 0.8789}
\definecolor{ieee-dark-dblue-100}{rgb}{0.0, 0.1562, 0.332}
\definecolor{ieee-dark-dblue-80}{rgb}{0.1797, 0.3008, 0.4609}
\definecolor{ieee-dark-dblue-60}{rgb}{0.3828, 0.4609, 0.5859}
\definecolor{ieee-dark-dblue-40}{rgb}{0.5781, 0.6289, 0.7188}
\definecolor{ieee-dark-dblue-20}{rgb}{0.7852, 0.8047, 0.8555}
\definecolor{ieee-dark-grey-100}{rgb}{0.457, 0.4688, 0.4805}
\definecolor{ieee-dark-grey-80}{rgb}{0.5625, 0.5625, 0.5742}
\definecolor{ieee-dark-grey-60}{rgb}{0.6641, 0.6641, 0.6758}
\definecolor{ieee-dark-grey-40}{rgb}{0.7734, 0.7695, 0.7773}
\definecolor{ieee-dark-grey-20}{rgb}{0.8789, 0.8828, 0.8828}
\definecolor{ieee-dark-black-100}{rgb}{0.0, 0.0, 0.0}
\definecolor{ieee-dark-black-80}{rgb}{0.3438, 0.3477, 0.3555}
\definecolor{ieee-dark-black-60}{rgb}{0.5, 0.5078, 0.5195}
\definecolor{ieee-dark-black-40}{rgb}{0.6523, 0.6602, 0.6719}
\definecolor{ieee-dark-black-20}{rgb}{0.8164, 0.8242, 0.8281}

\IfFileExists{identifier.tex}{
    \def\reviewpass{v3.0.0-06704da83fa48909-dirty-(490d2e1cb96474d687ba889e78012dc9c820ea3c)}

}{
    \def\reviewpass{Overleaf Version - Do not distribute!}
}

\newcommand{\includegraphicssafe}[2][]{\IfFileExists{#2}{
        \todo{#2}
    }{ 
        \todo{fig_src_latexpp/misc/not-found.jpg}
    }
}

\renewcommand{\arraystretch}{1.3}
\def\thetitle{A High-performance, Energy-efficient\\ Modular DMA Engine Architecture}

\newacronym{asic}{ASIC}{application-specific integrated circuit}
\newacronym{axi}{AXI}{advanced eXtensible interface}
\newacronym{bmc}{BMC}{baseboard management controller}
\newacronym{cpu}{CPU}{central processing unit}
\newacronym{dma}{DMA}{direct memory access}
\newacronym{dmac}{DMAC}{direct memory access controller}
\newacronym{dmae}{DMAE}{direct memory access engine}
\newacronym{dram}{DRAM}{dynamic random access memory}
\newacronym{dsp}{DSP}{digital signal processing}
\newacronym{dvfs}{DVFS}{dynamic voltage and frequency scaling}
\newacronym{e2e}{E2E}{end-to-end}
\newacronym{fifo}{FIFO}{first in, first out}
\newacronym{fpga}{FPGA}{field programmable gate array}
\newacronym{fpu}{FPU}{floating-point unit}
\newacronym{gemm}{GEMM}{general matrix multiply}
\newacronym{gp}{GP}{general-purpose}
\newacronym{hbm}{HBM}{high-bandwidth memory}
\newacronym{hil}{HIL}{hardware-in-the-loop}
\newacronym{hlc}{HLC}{high-level controller}
\newacronym{hpc}{HPC}{high-performance computing}
\newacronym{idma}{iDMA}{intelligent DMA}
\newacronym{ihls}{iHLS}{IP-based high-level synthesis}
\newacronym{ip}{IP}{intellectual property}
\newacronym{isa}{ISA}{instruction set architecture}
\newacronym{l1}{L1}{level-one}
\newacronym{l2}{L2}{level-two}
\newacronym{l3}{L3}{level-three}
\newacronym{llc}{LLC}{low-level controller}
\newacronym{mcu}{MCU}{microcontroller unit}
\newacronym{mimo}{MIMO}{multi-input multi-output}
\newacronym{ml}{ML}{machine learning}
\newacronym{noc}{NoC}{network-on-chip}
\newacronym{ooc}{OOC}{out-of-context}
\newacronym{obi}{OBI}{open bus protocol}
\newacronym{os}{OS}{operating system}
\newacronym{pcf}{PCF}{power control firmware}
\newacronym{pcs}{PCS}{power controller system}
\newacronym{pdk}{PDK}{process design kit}
\newacronym{pe}{PE}{processing element}
\newacronym{pfct}{PFCT}{periodic frequency control task}
\newacronym{pulp}{PULP}{parallel ultra-low power}
\newacronym{pvct}{PVCT}{periodic voltage control task}
\newacronym{pvt}{PVT}{process-voltage-temperature}
\newacronym{rpc}{RPC}{reduced pin count}
\newacronym{rtl}{RTL}{register transfer level}
\newacronym{sdma}{sDMAE}{sensor DMAE}
\newacronym{soa}{SoA}{state-of-the-art}
\newacronym[firstplural=systems on chip (SoCs)]{soc}{SoC}{system on chip}
\newacronym{spm}{SPM}{scratchpad memory}
\newacronym{spmv}{SpMV}{sparse matrix-vector multiply}
\newacronym{spmm}{SpMM}{sparse matrix-matrix multiply}
\newacronym{sram}{SRAM}{static read-only memory}
\newacronym{tcdm}{TCDM}{tightly-coupled data memory}
\newacronym{ulp}{ULP}{ultra-low-power}
\newacronym{vrm}{VRM}{voltage regulator module}

\newcommand{\gfs}{{GlobalFoundries'}}
\newcommand{\gftech}{{GF12LP+}}
\newcommand{\dc}{{Synopsys Design Compiler 2022.12}}

\newcommand{\occamy}{{Manticore-0432x2}}
\newcommand{\occamyfn}{\footnote{%
    {\occamy} is an adapted from the original Manticore architecture~\cite{9220474}; %
    Manticore-\emph{AAAA}x\emph{B}, where \emph{AAAA} represents the total number of \glspl{pe} and
    \emph{B} the number of chiplets in the system.%
}}
\newcommand{\cheshire}{{Cheshire}}
\newcommand{\mempool}{{MemPool}}
\newcommand{\controlpulp}{{ControlPULP}}
\newcommand{\pulpopen}{{PULP-open}}

\ifx\showacrcols\undefined
    \newcommand{\idma}{\gls{idma}}
    \newcommand{\idmae}{iDMAE}
    \newcommand{\idmaes}{iDMAEs}
    \newcommand{\dma}{\gls{dma}}
    
    \newcommand{\dmaes}{\glspl{dmae}}
    \newcommand{\dmae}{\gls{dmae}}

    \newcommand{\dmaa}{\gls{dmae} architecture}
\else
    \newcommand{\idma}{\textcolor{ieee-bright-red-100}{\gls{idma}}}
    \newcommand{\idmae}{\textcolor{ieee-dark-purple-100}{iDMAE}}
    \newcommand{\idmaes}{\textcolor{ieee-dark-lgreen-100}{iDMAEs}}
    \newcommand{\dma}{\textcolor{ieee-bright-orange-100}{\gls{dma}}}
    
    \newcommand{\dmaes}{\textcolor{ieee-bright-lgreen-100}{\glspl{dmae}}}
    \newcommand{\dmae}{\textcolor{ieee-bright-dgreen-100}{\gls{dmae}}}

    \newcommand{\dmaa}{\textcolor{ieee-bright-cyan-100}{\gls{dmae} architecture}}
\fi

\newcommand{\fe}{front-end}
\newcommand{\fes}{{\fe}s}
\newcommand{\me}{mid-end}
\newcommand{\mes}{{\me}s}
\newcommand{\be}{back-end}
\newcommand{\bes}{{\be}s}

\newcommand{\Fe}{Front-end}
\newcommand{\Fes}{Front-ends}
\newcommand{\Me}{Mid-end}
\newcommand{\Mes}{Mid-ends}
\newcommand{\Be}{Back-end}

\newcommand{\tus}{{\textunderscore}}

\begin{document}

\AddToShipoutPictureBG*{%
  \AtPageUpperLeft{%
    \hspace{\paperwidth}%
    \raisebox{-\baselineskip}{%
      \makebox[-35pt][r]{\footnotesize{
        \copyright~2023~IEEE. Personal use of this material is permitted. %
        Permission from IEEE must be obtained for all other uses, in any current or future media, including
      }}
}}}%

\AddToShipoutPictureBG*{%
  \AtPageUpperLeft{%
    \hspace{\paperwidth}%
    \raisebox{-2\baselineskip}{%
      \makebox[-37pt][r]{\footnotesize{
        reprinting/republishing this material for advertising or promotional purposes, creating new collective works, for resale or redistribution to servers or lists, or
      }}
}}}%

\AddToShipoutPictureBG*{%
  \AtPageUpperLeft{%
    \hspace{\paperwidth}%
    \raisebox{-3\baselineskip}{%
      \makebox[-185pt][r]{\footnotesize{
       reuse of any copyrighted component of this work in other works.
      }}
}}}%

\title{\thetitle}
\ifx\showrevision\undefined
    \newcommand{\todo}[1]{{#1}}
\else
    \newcommand{\todo}[1]{{\textcolor{red}{#1}}}
    \AddToShipoutPictureFG{%
        \put(%
            8mm,%
            \paperheight-1.5cm%
            ){\vtop{{\null}\makebox[0pt][c]{%
                \rotatebox[origin=c]{90}{%
                    \huge\textcolor{red!75}{\reviewpass}%
                }%
            }}%
        }%
    }
    \AddToShipoutPictureFG{%
        \put(%
            \paperwidth-6mm,%
            \paperheight-1.5cm%
            ){\vtop{{\null}\makebox[0pt][c]{%
                \rotatebox[origin=c]{90}{%
                    \huge\textcolor{red!30}{ETH Zurich - Unpublished - Confidential - Draft - Copyright Thomas Benz 2023}%
                }%
            }}%
        }%
    }
\fi

\ifx\showrebuttal\undefined
    \newcommand{\rev}[1]{#1}
\else
    \newcommand{\rev}[1]{{\textcolor{ieee-bright-lblue-100}{#1}}}
\fi

\ifx\showrebuttal\undefined
    \newcommand{\revdel}[1]{}
\else
    \newcommand{\revdel}[1]{\textcolor{ieee-bright-red-100}{\st{#1}}}
\fi

\ifx\showrebuttal\undefined
    \newcommand{\revrep}[2]{#2}
\else
    \newcommand{\revrep}[2]{\revdel{#1} \rev{#2}}
\fi

\ifx\showrebuttal\undefined
    \newcommand{\revprg}[1]{}
\else
    \newcommand{\revprg}[1]{\hspace{-0.5ex}\textcolor{red}{\scalebox{.2}[1.5]{$\blacksquare$}}\hspace{-0.5ex}}
\fi

\author{Thomas~Benz\orcidlink{0000-0002-0326-9676},~\IEEEmembership{Student Member, IEEE},
        Michael~Rogenmoser\orcidlink{0000-0003-4622-4862},~\IEEEmembership{Student Member, IEEE},
        Paul~Scheffler\orcidlink{0000-0003-4230-1381},~\IEEEmembership{Student Member, IEEE},
        Samuel~Riedel\orcidlink{0000-0002-5772-6377},~\IEEEmembership{Student Member, IEEE},
        Alessandro~Ottaviano\orcidlink{0009-0000-9924-3536},~\IEEEmembership{Student Member, IEEE},
        Andreas~Kurth\orcidlink{0000-0001-5613-9544},~\IEEEmembership{Member, IEEE},
        Torsten~Hoefler\orcidlink{0000-0002-1333-9797},~\IEEEmembership{Fellow,~IEEE},
        and~Luca~Benini\orcidlink{0000-0001-8068-3806},~\IEEEmembership{Fellow, IEEE}%
        \IEEEcompsocitemizethanks{\IEEEcompsocthanksitem T.~Benz, M.~Rogenmoser, P.~Scheffler, S.~Riedel, A.~Ottaviano, A.~Kurth and L.~Benini are with the Integrated Systems Laboratory (IIS), ETH Zurich, Switzerland\protect\\
        E-mail: \{tbenz,michaero,paulsc,sriedel,aottaviano,akurth,lbenini\}@ethz.ch
        \IEEEcompsocthanksitem T.~Hoefler is with the Scalable Parallel Computing Laboratory (SPCL), ETH Zurich. E-mail: htor@inf.ethz.ch
        \IEEEcompsocthanksitem L.~Benini also is with Department of Electrical, Electronic and Information Engineering (DEI), University of Bologna, Bologna, Italy.}%
}

\IEEEtitleabstractindextext{%
\begin{abstract}

Data transfers are essential in today’s computing systems as latency and complex memory access patterns are increasingly challenging to manage. %
\Glsplf{dmae} are critically needed to transfer data independently of the processing elements, hiding latency and achieving high throughput even for complex access patterns to high-latency memory. %
With the prevalence of heterogeneous systems, {\dmaes} must operate efficiently in increasingly diverse environments. %
This work proposes a modular and highly configurable open-source {\dmaa} called {\glsf{idma}}, split %
into three parts that can be composed and customized independently. %
The {\fe} implements the control plane binding to the surrounding system. %
The {\me} accelerates complex data transfer patterns such as multi-dimensional transfers, scattering, or gathering. %
The {\be} interfaces with the on-chip communication fabric (data plane). %
We assess the efficiency of {\idma} in various instantiations: %
In high-performance systems, we achieve speedups of up to 15.8\x~with only \SI{1}{\percent} additional area compared to a base system without a {\dmae}. %
We achieve an area reduction of \SI{10}{\percent} while improving ML inference performance by \SI{23}{\percent} in ultra-low-energy edge AI systems over an existing {\dmae} solution. %
We provide area, timing, latency, and performance characterization to guide its instantiation in various systems. %
\end{abstract}

\begin{IEEEkeywords}
DMA, DMAC, Direct Memory Access, Memory Systems, High-performance, Energy-efficiency, Edge AI, AXI, TileLink
\end{IEEEkeywords}
}

\glsresetall

\maketitle
\IEEEdisplaynontitleabstractindextext

\IEEEraisesectionheading{%
    \section{Introduction}%
    \label{sec:introduction}%
}

\IEEEPARstart{D}{irect} memory access engines (DMAEs) form the communication backbone of many contemporary computers~\cite{ma_mt-dma_2019}. %
They concurrently move data at high throughput while hiding memory latency and minimizing processor load, freeing the latter to do useful compute. %
This function becomes increasingly critical with the trend towards physically larger systems~\cite{9361255} and ever-increasing memory bandwidths~\cite{9893362}. %
With Moore's Law slowing down, 2.5D and 3D integration are required to satisfy future applications' computational and memory needs, leading to wider and higher-bandwidth memory systems and longer access latencies~\cite{9567038, 9381710}. %

Without {\dmaes}, \glspl{pe} need to read and write data from and to remote memory, often relying on deep cache hierarchies to mitigate performance and energy overheads. %
This paper focuses on explicitly managed memory hierarchies, where copies across the hierarchy are handled by {\dmaes}. %
We refer the interested reader to \cite{10.1145/3534962, nagarajan2020primer, jain2019cache, balasubramonian2011multi} for excellent surveys on cache-based memory systems. %
Caches and {\dmaes} often coexist in modern computing systems as they address different application needs. %
Dedicated {\dmaes} are introduced to efficiently and autonomously move data for workloads where memory access is predictable, weakly data-dependent, and made in fairly large chunks, decoupling memory accesses from execution and helping maximize PE time spent on useful compute. %

When integrating {\dmaes}, three main design challenges must be tackled: the control-plane interface to the \glspl{pe}, the intrinsic data movement capabilities of the engine, and the on-chip protocols supported in the data plane. %
The sheer number of {\dmaes} present in literature and available as commercial products explains why these choices are usually fixed at design time. %
The increased heterogeneity in today's accelerator-rich computing environments leads to even more diverse requirements for {\dmaes}. %
Different on-chip protocols, programming models, and application profiles lead to a large variety of different {\glsf{dma}} units used in modern \glspl{soc}, hindering integration and verification efforts. %

We present a modular and highly parametric {\dmaa} called \emph{\glsf{idma}}, \revrep{that}{which} is composed of three distinct parts: the \emph{\fe} handling \gls{pe} interaction, the \emph{\me} managing the engine's lower-level data movement capabilities, and the \emph{\be} implementing one or more on-chip protocol interfaces. %
We call concrete implementations of \revrep{or}{our} {\idma} architecture \emph{\idmaes}.
All module boundaries are standardized to facilitate the substitution of individual parts, allowing for the same {\dmaa} to be used across a wide range of systems and applications. %
The synthesizable \gls{rtl} description of {\idma}, silicon-proven in various instances, and the system bindings are available free and open-source under a libre Apache-based license\footnote{\texttt{\url{https://github.com/pulp-platform/iDMA}} for {\idma}, system repositories in the same group (e.g., \texttt{mempool}) for integrations.}.

In more detail, this paper makes the following contributions: %
\begin{itemize}
    \item We specify a \emph{modular}, \emph{parametric} {\dmaa} composed of interchangeable parts, allowing {\idma} to accommodate and benefit any system. %
    \item We optimize {\idma} to minimize hardware buffering through a highly agile, read-write decoupled, dataflow-oriented transport engine that maximizes bus utilization in any context.  %
    Our architecture incurs no idle time between transactions, even when adapting between different on-chip protocols, and incurs only \emph{two} cycles of initial latency to launch a multi-dimensional affine transfer. %
    \item We propose and implement a two-stage transfer acceleration scheme: the {\me}\rev{s} manage\revdel{s} (distribute\revdel{s},  repeat\revdel{s}, and modif\revrep{ies}{y}) transfers while an in-stream \emph{acceleration port} enables configurable in-flight operation on the data being transferred. %
    \item We present and implement multiple system bindings ({\fes}) and industry-standard on-chip protocols ({\bes}), allowing our engines to be used in a wide range of contexts\rev{,} from \gls{ulp} to \gls{hpc} systems. %
    A lightweight data initialization feature allows {\idma} to initialize memory given various data patterns. %
    \item We thoroughly characterize our architecture in area, timing, and latency by creating area and timing models with less than \SI{9}{\percent} mean error and an analytical latency model, easing instantiation in third-party designs and accelerating system prototyping. %
    \item We use synthetic workloads to show that {\idma} achieves high bus utilization in ultra-deep memory systems with affordable area growth. %
    Our architecture perfectly hides latency in systems with memory hierarchies hundreds of stages deep. %
    It reaches full bus utilization on transfers as small as \SI{16}{\byte} while occupying an area footprint of less than \SI{25}{\kilo\gateeq}~%
    \rev{\footnote{%
    \rev{Gate equivalent (\si{\gateeq}) is a technology-independent figure of merit measuring circuit complexity. %
    A \si{\gateeq} represents the area of a two-input, minimum-strength {NAND} gate.} %
    }%
    } in a 32-\si{\bit} configuration. %
\end{itemize}

Furthermore, we conduct five system integration studies implementing and evaluating {\idma} in systems spanning a wide range of performance and complexity: %
\begin{itemize}
    \item On a minimal single-core, Linux-capable \gls{soc}~\cite{ottaviano2023cheshire} with reduced pin count \glsunset{dram}\gls{dram}, our {\idmae} reaches speedups of up to 6\x~on fine-granular \emph{64}-\si{\byte} transfers over an off-the-shelf \gls{dma} while reducing \gls{fpga} resource requirements by more than \SI{10}{\percent}. %
    \item In \revrep{an}{a} \gls{ulp} multicore edge node~\cite{rossi2014ultra} for edge AI applications, we replace the cluster {\dma} with our {\idmae} and improve the inference performance of {MobileNetV1} from {7.9}~{MAC/cycle} to {8.3}~{MAC/cycle} while reducing \gls{dmae} area by \SI{10}{\percent}. %
    \item We add our {\idmae} to a real-time system~\cite{Ottaviano2023ControlPULPAR} and implement a dedicated {\me} autonomously launching repeated 3D transfer tasks to reduce core load. %
    Our {\me} incurs an area penalty of \SI{11}{\kilo\gateeq}, negligible compared to the surrounding system. %
    \item In a scaled-out cluster manycore\revprg{~\cite{Riedel2023}} architecture\rev{~\cite{Riedel2023}}, we integrate a novel {distributed \revrep{\dmae}{\idmae}}~\cite{Riedel2023} with multiple {\bes}, accelerating integer workloads by 15.8\x~while increasing the area by less than \SI{1}{\percent}. %
    \item In a dual-chiplet multi-cluster manycore based on Manticore~\cite{9220474}, adding {\idmae} to each compute cluster enables speedups of up to 1.5\x~and 8.4\x~on dense and sparse floating-point workloads, respectively, while incurring only \SI{2.1}{\percent} in cluster area compared to a baseline architecture without a {\dma}. %
\end{itemize}

\section{Architecture}
\label{sec:arch}

\begin{figure}
    \centering%
    \includegraphics[width=\columnwidth]{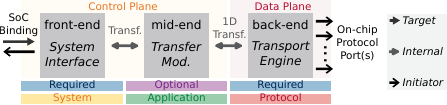}%
    \caption{%
        Schematic of {\idma}: Our engines are split into three parts: at least one {\fe}, one or multiple optional {\mes}, and at least one {\be}. %
    }%
    \label{fig:arch:arch}%
\end{figure}

Unlike \glsf{soa} {\dmaes}, we propose a modular and highly parametric \revrep{\dma}{\dmae} architecture composed of three distinct parts, as shown in \Cref{fig:arch:arch}. %
The \emph{\fe} defines the interface through which the processor cores control the {\dmae}, corresponding to the \emph{control plane}. %
The \emph{\be} or \emph{data plane} implements the on-chip network manager port(s) through which the {\dmae} moves data. %
Complex and capable on-chip protocols like \gls{axi}~\cite{arm_amba_2021}, to move data through the system\rev{,} and simpler core-local protocols like \gls{obi}~\cite{obi_spec_nodate} to connect to \gls{pe}-local memories, are supported by the {\be}. %
The \emph{\me}, connecting the front- and {\be}, slices complex transfer descriptors provided by the {\fe} (e.g., when transferring N-dimensional tensors) into one or multiple simple 1D transfer descriptors for the {\be} to process.
In addition, multiple {\mes} may be chained to enable complex transfer processing steps. %
Our {\controlpulp} case study in \Cref{sec:sys:control-pulp} shows this chaining mechanism by connecting a real-time and a {3D tensor} {\me} to efficiently address that platform's needs. %

\begin{figure}
    \centering%
    \includegraphics[width=\columnwidth]{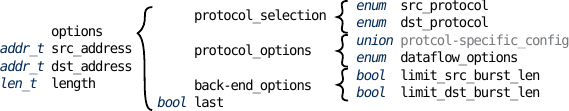}%
    \caption{%
        Outline of the \emph{1D transfer descriptor} (exchanged between mid- and {\be}).%
    }%
    \label{fig:arch:arch-request}%
\end{figure}

To ensure compatibility between these three different parts, we specify their interfaces.
From the {\fe} or the last {\me}, the {\be} accepts a \emph{1D transfer descriptor} specifying a \emph{source address}, a \emph{destination address}, \emph{transfer length}, \emph{protocol}, and \emph{{\be} options}, as seen in \Cref{fig:arch:arch-request}. %
{\Mes} receive bundles of {\me} configuration information and a \emph{1D transfer descriptor}. %
A {\me} will strip its configuration information while modifying the \emph{1D transfer descriptor}. %
\rev{%
All interfaces between front-, mid-, and {\bes} feature \emph{ready-valid} handshaking and can thus be pipelined.
}

\subsection{\Fe}
We present three {\fe} types: a simple and area-efficient register-based scheme, an efficient microcode programming interface, and a high-performance descriptor-based {\fe}, as shown in \cref{tab:arch:fe}. %
Our selection is tailored to the current set of use cases; different {\fes} can easily be created, e.g., allowing us to use our descriptor-based binding in 32-\si{\bit} systems. %

\begin{table}
    \centering
    \caption{%
        Identifiers and descriptions of {\fes} employed in the use cases. %
        {\Fes} in \textcolor{ieee-dark-grey-100}{gray} are available but not further discussed in this work. %
    }%
    \renewcommand*{\arraystretch}{1.0}
    \begin{tabular}{@{}lll@{}}\toprule
        \textbf{\Fe} &
        \textbf{Description} &
        \textbf{Conf.} \\ \midrule
        \textcolor{ieee-dark-grey-100}{\emph{reg\tus32}} &
        \multirow{5}{*}{%
            \makecell[tl]{%
                Core-private register-based \\
                configuration interface for \gls{ulp}-systems%
            }%
        } &
        \textcolor{ieee-dark-grey-100}{32-\si{\bit}, 1D} \\
        \textcolor{ieee-dark-grey-100}{\emph{reg\tus32\tus2d}} &                                                              &
        \textcolor{ieee-dark-grey-100}{32-\si{\bit}, 2D} \\
        \emph{reg\tus32\tus3d} &
        & 
        32-\si{\bit}, 3D \\
        \textcolor{ieee-dark-grey-100}{\emph{reg\tus64}} &
        &
        \textcolor{ieee-dark-grey-100}{64-\si{\bit}, 1D} \\
        \textcolor{ieee-dark-grey-100}{\rev{\emph{reg\tus64\tus2d}}} &
        &
        \textcolor{ieee-dark-grey-100}{\rev{64-\si{\bit}, 2D}} \\%
        \arrayrulecolor{ieee-dark-grey-100}\midrule
        \emph{reg\tus32\tus{rt}\tus3d}&
        \makecell[tl]{%
            Core-private register-based system \\
            binding supporting our real-time {\me}%
        } %
        &
        32-\si{\bit}, 3D \\
        \arrayrulecolor{ieee-dark-grey-100}\midrule
        \emph{desc\tus64} &
        \makecell[tl]{%
            Transfer-descriptor-based interface \\%
            designed for 64-\si{\bit} systems compatible \\%
            with the Linux \gls{dma} interface%
        }&
        64-bit, 1D \\
        \arrayrulecolor{ieee-dark-grey-100}\midrule
        \emph{inst\tus64} &
        \makecell[tl]{%
            Interface decoding a custom \revrep{\dma}{\idma}\\%
            \emph{RISC-V} instructions used in \gls{hpc} systems%
        } &
        64-\si{\bit}, 2D \\
        \arrayrulecolor{black}\bottomrule
    \end{tabular}
    \label{tab:arch:fe}
\end{table}

\paragraph*{\textbf{Register-based}}
Core-private register-based configuration interfaces are the simplest {\fes}. %
Each \gls{pe} uses its own dedicated configuration space to eliminate race conditions while programming the {\dmae}~\cite{rossi2014ultra}. %
We employ different memory-mapped register layouts depending on the host system's word width and whether a \emph{multi-dimensional tensor \me} is present. %
The \emph{src\_address}, \emph{dst\_address}, \emph{transfer\_length}, \emph{status}, \emph{configuration}, and \emph{transfer\_id} registers are shared between all variants. %
In the case of multi-dimensional configuration, every tensor dimension introduces three additional fields: \emph{src\_stride}, \emph{dst\_stride}, and \emph{num\_repetitions}. %
After configuring the shape of a transfer, it is launched by reading from  \emph{transfer\_id}, which returns an incrementing unique transfer ID. %
The \rev{ID} last completed \revdel{ID} may be read from the \emph{status} register, enabling transfer-level synchronization. %

\paragraph*{\textbf{Descriptor-based}}
As the use of \emph{transfer descriptors} is common practice\revprg{~\cite{ma_mt-dma_2019, synopsys_designware_nodate}} in Linux-capable multicore systems\rev{~\cite{ma_mt-dma_2019, synopsys_designware_nodate}}, we provide \emph{desc\tus64}, a 64-bit {\fe} compatible with the \emph{Linux {\dma}} interface. %
Given a pointer, the {\fe} uses a dedicated manager port to fetch transfer descriptors from memory; currently, the AXI, AXI-Lite~\cite{arm_amba_2021}, and OBI~\cite{obi_spec_nodate} protocols are supported. %
The descriptors consist of a \emph{src\_address}, \emph{dst\_address}, \emph{transfer\_length}, and a run-time \emph{backend\_configuration}, corresponding to information required for a \textit{1D} transfer. %
Descriptor chaining~\cite{ma_mt-dma_2019} is supported to allow efficient long and arbitrarily shaped transfers. %

\paragraph*{\textbf{Instruction-based}}
We present \emph{inst\tus64}, a {\fe} that can be tightly coupled to a RISC-V core encoding \revrep{\dma}{\idma} transfers directly as instructions. %
For example, a \emph{Snitch}~\cite{Zaruba_2021} RISC-V core using \emph{inst\tus64} can launch a transaction within three cycles, enabling highly agile data transfers. %

\subsection{\Me}

In {\idma}, {\mes} process complex transfers coming from the {\fe} and decompose them into one or multiple 1D transfer(s), which can \revrep{directly be handled}{be handled directly} by the {\be}. %
Our \revrep{mid-ends}{\me} overview can be found in \cref{tab:arch:me}.

\begin{table}
    \centering
    \caption{%
        Identifier of implemented {\mes}. %
    }%
    \begin{tabular}{@{}ll@{}}\toprule
        \textbf{\Me} &
        \textbf{Description} \\
        \midrule
        \emph{tensor\tus2D} &
        \makecell[tl]{Optimized to accelerate 2D transfers.} \\
        \emph{tensor{\tus}ND} &
        \makecell[tl]{{\Me} accelerating ND transfers.} \\
        \emph{mp{\tus}split} &
        \makecell[tl]{%
            {\Me} splitting transfers along a parametric \\%
            address boundary.%
        } \\
        \emph{mp{\tus}dist} &
        \makecell[tl]{%
            {\Me} distributing transfers over multiple \bes.} \\
        \emph{rt{\tus}3D} &
        \makecell[tl]{%
            {\Me} repetitively launching 3D transfers designed \\
            for real-time systems.%
        } \\
        \arrayrulecolor{black}\bottomrule
    \end{tabular}
    \label{tab:arch:me}
\end{table}

\paragraph*{\textbf{Tensor \Mes}}

To support multi-dimensional transfers, two distinct {\mes} are provided. %
\emph{Tensor\tus2D} supports 2D transfers through an interface common in embedded systems which requires the source and destination strides, the total length of the transfer, the base address, and the length of the 1D transfers it is composed of. %
\emph{Tensor{\tus}ND} can be parameterized at compile-time to support transfers on tensors of any dimension \emph{N}. %
It is programmed by providing the source address, destination address, number of repetitions, and strides for each dimension. %

\paragraph*{\textbf{Distribution \Me}}

In large manycore systems like {\mempool}~\cite{Riedel2023}, one centralized actor may want to schedule the data requests of multiple interconnect manager ports. We implement this functionality with a \rev{distributed} \emph{multi-\be} {\idmae}. %
To distribute work among {\bes}, we create two specialized \mes{} called \emph{mp{\tus}split} and \emph{mp{\tus}dist}. %
\emph{Mp{\tus}split} splits a single linear transfer into multiple transfers aligned to a parametric address boundary, guaranteeing that no resulting transfer crosses specific address boundaries. %
Which is required when sending distributed transfers to multiple {\bes}, see \Cref{sec:sys:mempool}. %
\emph{Mp{\tus}dist} then distributes the split transfers over multiple parallel downstream mid- or \bes{}, arbitrating the transfers based on their address offsets. %
\emph{Mp{\tus}dist}s' number of outgoing ports is set per default to two. %

\paragraph*{\textbf{Real-time \Me}}
\label{subsubsec:midend-controlpulp}

Repeated ND transfers are often required for data acquisition tasks such as reading out sensor arrays in real-time systems featuring complex address maps. %
The number of dimensions, \emph{N}, can be set at compile time. To relieve the pressure on general-purpose cores already involved in task scheduling and computation, this can be done by \revrep{our iDMAE}{\idma} using a specialized {\me}. %
The \emph{rt{\tus}3D} {\me} enables a configurable number of repeated 3D transactions whose periodicity and transfer shape \revrep{is}{are} configured via the {\fe}. %
A bypass mechanism allows the core to dispatch unrelated transfers using the same \revrep{front-end and iDMAE back-end}{front- and {\be}}. %

\subsection{\Be}

\begin{figure}
    \centering%
    \includegraphics[width=\columnwidth]{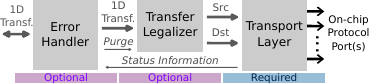}%
    \caption{%
        The internal architecture of the {\be}. %
        The \emph{transport layer} handles the actual copying of the data on the on-chip protocol, supported by the optional \emph{transfer legalizer} and the \emph{error handler}.
    }%
    \label{fig:arch:arch-backend}%
\end{figure}

For given on-chip protocols, a {\be} implements efficient \emph{in-order} \emph{one-dimensional arbitrary-length} transfers. %
\emph{\revrep{Out-of-order}{Multichannel}} {\dmaes}\rev{, featuring multiple ports into the memory system,} \revrep{are}{can be} built by connecting multiple \revdel{in-order }{\bes} to a single front- or {\me}, requiring \rev{a} similar area as \revrep{an out-of-order}{true multichannel} {\be}\revdel{,} but\revdel{ much} less \rev{verification and design} complexity. %
\rev{%
Arbitration between the individual {\bes} can either be done explicitly by choosing the executing {\be} through software or by an \emph{arbitration} {\me} using round-robin or address-based distribution schemes. %
Similar hardware will be required in a true multichannel {\be} to distribute transactions to available channels. %
}

The {\be} comprises three parts, see \cref{fig:arch:arch-backend}. The \emph{error handler} communicates and reacts to failing transfers, the \emph{transfer legalizer} reshapes incoming transfers to meet protocol requirements, and the \emph{transport layer} handles data movement and possibly in-cycle switches between protocol-specific data plane ports. %
Of these three units, only the transport layer is mandatory. %

\paragraph*{\textbf{Error Handler}} 

An \emph{error handler} may be included if the system or application requires protocol error reporting or handling. %
Our current error handler can either \emph{continue}, \emph{abort}, or \emph{replay} erroneous transfers. %
Replaying erroneous transfers allows complex ND transfers to continue in case of errors in single {\be} iterations without the need to abort and restart the entire transfer. %

When an error occurs, the {\be} pauses the transfer processing and passes the offending transfer's legalized burst base address to its {\fe}.  %
The \glspl{pe} can then specify through the {\fe} which of the three possible actions the error handler should take \rev{to resolve the situation}. %

\paragraph*{\textbf{Transfer Legalizer}}

\begin{figure}[t]
    \centering%
    \includegraphics[width=\columnwidth]{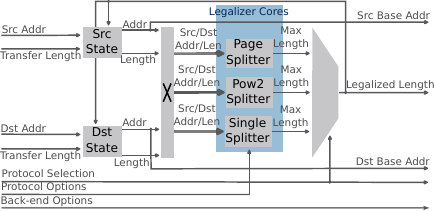}%
    \caption{%
        \rev{%
        The internal architecture of the \emph{transfer legalizer}. %
        Any given transfer can be legalized except \rev{for} zero-length transactions: They may optionally be rejected. %
        }%
    }%
    \label{fig:arch:arch-legalizer}%
\end{figure}

Shown in  \Cref{fig:arch:arch-legalizer}, it accepts a 1D transfer and legalizes it to be supported by the specific on-chip protocol(s) in use. %
Transfer information is stored internally and modular \emph{legalizer cores} determine the \rev{transfer's} maximum legal length\revdel{ of transfer} supported given user constraints and the protocols' properties. %
For protocols that do not support bursts, the legalizer decomposes transfers into individual bus-sized accesses. %
Otherwise, splitting happens at page boundaries, the maximum burst length supported by the protocol, or user-specified burst length limitations. %
The source and destination protocols' requirements are considered to guarantee only legal transfers are emitted. %
In area-constrained designs, the transfer legalizer may be omitted; legal transfers must be guaranteed in software. %

\paragraph*{\textbf{Transport Layer}}
\label{sec:arch_transport-layer}

\begin{figure}
    \centering%
    \includegraphics[width=\columnwidth]{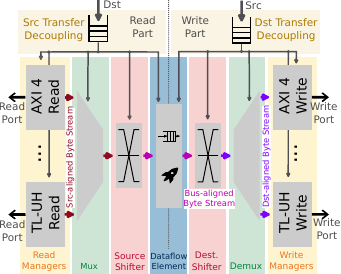}%
    \caption{%
        The architecture of the \emph{transport layer}. %
        One or multiple \emph{read manager(s)} feed a stream of bytes into the \emph{source shifter}, the \emph{data flow element}, the \emph{destination shifter}, and finally, into one or multiple \emph{write manager(s)}. %
        \rev{\inlinegraphicscap{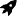} denotes the \emph{in-stream accelerator}.} %
    }%
    \label{fig:arch:arch-transport}%
\end{figure}

The parametric and modular \emph{transport layer} implements the protocol characteristics for legalized transfers and decouples read and write operations, \rev{maximizing the bus utilization of any transfers.} %
It uses \emph{read} and \emph{write managers} to handle protocol-specific operations, allowing it to \rev{internally} operate\revdel{ internally} only on generic byte streams, as shown in \Cref{fig:arch:arch-transport}. %
This enables our \revrep{DMAE}{\idma} to easily support multiple on-chip protocols and multiple ports of the same protocol. %
The number and type of protocol ports available in the engine must be set at compile time, whereas the given protocol port a transaction uses can be selected during run-time through the {\fe}. %

The read and write parts of the transport layer are decoupled from the legalizer by \gls{fifo} buffers, allowing a configurable number of outstanding transfers. %
A \emph{dataflow element} decouples the read and write parts, ensuring that only protocol-legal back pressure is applied to the memory system at each end, \rev{coalesces transfers}, and cuts long timing paths to increase the engine's maximum operating frequency.
Fully buffered operation may be required depending on the system and the memory endpoints; in this case, the small \gls{fifo} buffer in the dataflow element may be replaced with an \glsunset{sram}\gls{sram}-based buffer, allowing entire transfers to be stored. %
Two shifters, one at each end of the dataflow element, align the byte stream to bus boundaries.
\emph{In-stream accelerators}, allowing operations performed on the data stream during data movement, may be integrated into the dataflow element, augmenting the buffer in the transport layer.
Our dataflow-oriented architecture allows \rev{us} to switch between multiple read managers, write managers, and in-stream accelerators \emph{in-cycle}, allowing our engine to asymptotically reach perfect bus utilization even when the used protocols or acceleration schemes change regularly. %

\paragraph*{\textbf{Protocol Managers}}
\label{sec:arch:protocol-managers}

\begin{table}
    \centering
    \scriptsize{%
    \caption{%
        Available on-chip protocols and their key characteristics. All protocols share \emph{byte addressability} and a \emph{ready-valid} handshake.%
    }%
    \label{tab:arch:managers}
    \renewcommand*{\arraystretch}{1.0}
    \begin{threeparttable}
        \addtolength{\tabcolsep}{1pt}
        \begin{tabular}{@{}lllllll@{}} \toprule
            \textbf{Protocol} &
            \textbf{Version} &
            \textbf{\makecell[cl]{Request \\ Channel}} &
            \textbf{\makecell[cl]{Response \\ Channel}} &
            \textbf{Bursts} \\
            \midrule
            \emph{AXI4+ATOP}~\cite{arm_amba_2021} &
            \emph{H.c} &
            \emph{AW~\tnote{b}, W~\tnote{b}, AR~\tnote{a}} &
            \emph{B~\tnote{b}, R~\tnote{a}} &
            \makecell[cl]{\emph{256 beats} \\ or \emph{\SI{4}{\kilo\byte}~\tnote{c}}} \\
            \emph{AXI4 Lite}~\cite{arm_amba_2021} &
            \emph{H.c} &
            \emph{AW~\tnote{b}, W~\tnote{b}, AR~\tnote{a}} &
            \emph{B~\tnote{b}, R~\tnote{a}} &
            no \\
            \emph{AXI4 Stream}~\cite{axi_stream_nodate} &
            \emph{B} &
            \emph{T~\tnote{d}} &
            \emph{T~\tnote{d}} &
            unlimited \\
            \emph{OpenHW OBI}~\cite{obi_spec_nodate} &
            \emph{v1.5.0} &
            \emph{D} &
            \emph{R} &
            no \\
            \emph{SiFive TileLink}~\cite{tl_spec_nodate}~\tnote{e} &
            \emph{v1.8.1} &
            \emph{A} &
            \emph{R} &
            \makecell[cl]{UH: \\ \emph{power of two}} \\
            \emph{Init}~\tnote{g}~\tnote{h} &
            N.A. &
            N.A. &
            N.A. &
            N.A. \\
            \bottomrule
        \end{tabular}
        \addtolength{\tabcolsep}{-3pt}
        \begin{tablenotes}[para, flushleft]
            \item[a] read
            \item[b] write
            \item[c] whichever is reached first
            \item[d] symmetrical \emph{RX/TX} channels
            \item[e] \emph{TL-UL} \& \emph{TL-UH} supported
            \item[f] \emph{valid} is expressed through \emph{cyc}, \emph{ready} through \emph{ack} (and \emph{rti})
            \item[g] memory initialization pseudo protocol
            \item[h] read-only supported
        \end{tablenotes}
    \end{threeparttable}
    }
\end{table}

The transport layer abstracts the interfacing on-chip protocols through \emph{read} and \emph{write managers} with standardized interfaces, allowing for true multi-protocol capabilities. %
\emph{Read managers} receive the current read transfer's base address, transfer length, and protocol-specific configuration information as input\rev{s}. %
They then emit a \emph{read-aligned} stream of data bytes to the downstream transport layer. %
\emph{Write managers} receive the write transfer information and the \emph{write-aligned} stream of data bytes from the upstream  transport layer to be emitted over their on-chip protocol's manager port. %

\Cref{tab:arch:managers} provides a complete list of supported protocols. %
The \emph{Init} pseudo-protocol only provides a \emph{read manager} emitting a configurable stream of either the \emph{same repeated value}, \emph{incrementing values}, or a \emph{pseudorandom sequence}. This enables our engine to accelerate memory initialization. %

\section{Case Studies}
\label{sec:case-studies}

To demonstrate the generality and real-world benefits of \revdel{our} {\idma}, we detail its integration into five systems spanning a wide range of capabilities, from \gls{ulp} processors for edge AI, to \gls{hpc} manycore architectures. %

\subsection{{\pulpopen}}
\label{sec:sys:pulp}

{\pulpopen} is a \gls{ulp} edge compute platform consisting of a 32-\si{\bit} {RISC-V} microcontroller host and a parallel compute cluster~\cite{pullini2019mr}. %
The compute cluster comprises eight 32-\si{\bit} {RISC-V} cores with custom \gls{isa} extensions to accelerate \gls{dsp} and \glsu{ml} workloads, enabling energy-efficient \gls{ml} inference in extreme-edge AI nodes. %
These cores are connected to an \gls{sram}-based \gls{tcdm} with single-cycle access latency, providing the processing cores with fast access to shared data. %
While the \gls{tcdm} is fast, it is very limited in size;
the platform thus features a \gls{l2} on-chip and \gls{l3} off-chip HyperBus RAM~\cite{hyperbus-datasheet}. %
To allow the cluster fast access to these larger memories, a {\dma} unit is embedded, specialized for transferring data from and to the \gls{l1} memory. %

\paragraph*{\textbf{{\idmae} Integration}}

\begin{figure}
    \centering%
        \subfloat{\includegraphics[width=1\columnwidth]{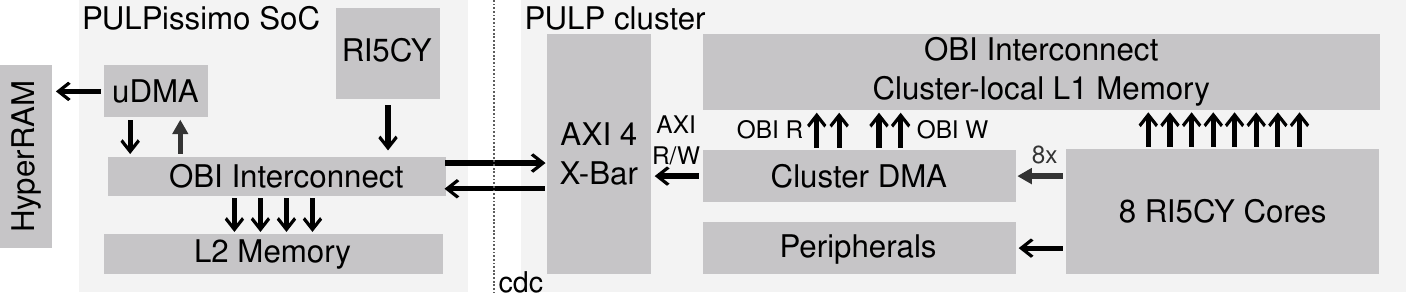}%
        \label{fig:case:pulp:bd}%
        }

        \subfloat{\includegraphics[width=0.9\columnwidth]{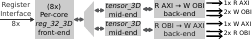}%
        \label{fig:case:pulp:dma}%
        }
    \caption{%
        (Top) Block diagram of the {\pulpopen} system. %
        (Bottom) Configuration of the cluster {\idmae}. %
    }%
    \label{fig:case:pulp:arch}%
\end{figure}

In the {\pulpopen} system, our {\idmae} is integrated into the processing cluster with a 64-\si{\bit} \emph{AXI4} interface to the host platform and an \emph{OBI} connection to the \gls{tcdm}, see \Cref{fig:case:pulp:arch}. %
The multi-protocol {\be} is fed by a \emph{tensor{\tus}ND} {\me}, configured to support three dimensions, allowing for fast transfer of 3D data structures common in \gls{ml} workloads. %
At the same time, higher-dimensional transfers are handled in software. %
The back- and {\me} are configured through per-core \emph{reg\_32\_3d} {\fes} and two additional {\fes}, allowing the host processor to configure the {\idmae}. %
\rev{%
Round-robin arbitration is implemented through a round-robin arbitration {\me} connecting the {\fes} to the \emph{tensor{\tus}ND} {\me}.%
} %
Multiple per-core {\fes} ensure atomic {\dma} access and prevent interference between the cores launching transactions.

\paragraph*{\textbf{Benchmarks}}

To evaluate {\idmae} performance in a realistic application, we use Dory~\cite{burrello2020dory} to implement {MobileNetV1} inference on {\pulpopen}. %
This workload relies heavily on the {\idmae} to transfer the data for each layer stored in \gls{l2} or off-chip in \gls{l3} into the cluster's \gls{tcdm} in parallel with cluster core computation. %
2D, 3D, and very small transfers are frequently required for this workload. %

In previous versions of {\pulpopen}, MCHAN~\cite{rossi2014ultra} was used to transfer data between the host \gls{l2} and the cluster's \gls{tcdm}. We assume this as a baseline for our evaluation. %

\paragraph*{\textbf{Results}}

In {\pulpopen}, {\idmae} can almost fully utilize the bandwidth to the \gls{l2} and \gls{tcdm} in both directions: %
measuring with the on-board timer, a transfer of \SI{8}{\kibi\byte} from the cluster's \gls{tcdm} to \gls{l2} requires \emph{1107} cycles, \revdel{where}{of which} \emph{1024} cycles are required to transfer the data using a 64\rev{-}\si{\bit} data bus. The minimal overhead is caused by configuration, system latency, and contention with other ongoing memory accesses. %
During {MobileNetV1} inference, individual cores frequently require short transfers, incurring a potentially high configuration overhead. %
With its improved \emph{tensor\_3D} {\me}, {\idma} improves the cores' utilization and throughput for the network over MCHAN, achieving an average of \SI{8.3}{MAC\per cycle} compared to the previously measured \SI{7.9}{MAC\per cycle}. %
Furthermore, configured with similar queue depths as MCHAN, {\idmae} with its \emph{reg\_32\_3d} achieves a \SI{10}{\percent} reduction in the utilized area within a PULP cluster. %

\subsection{{\controlpulp}}
\label{sec:sys:control-pulp}

{\controlpulp}~\cite{Ottaviano2023ControlPULPAR} is an on-chip parallel \gls{mcu} employed as a \gls{pcs} for manycore \gls{hpc} processors. %
It comprises a single 32-\si{\bit} {RISC-V} \emph{manager domain} with \SI{512}{\kibi\byte} of \gls{l2} scratchpad memory and a programmable accelerator (\emph{cluster domain}) hosting eight 32-\si{\bit} RISC-V cores and \SI{128}{\kibi\byte} of \gls{tcdm}. %

A \gls{pcf} running on FreeRTOS implements a reactive power management policy. %
{\controlpulp} receives (i) \gls{dvfs} directives such as frequency target and power budget from high-level controllers and (ii) temperature from \gls{pvt} sensors and power consumption from \glspl{vrm}, and is tasked to meet its constraints. %
The \gls{pcf} consists of two periodic tasks, \gls{pfct} (low priority) and \gls{pvct} (high priority) that handle the power management policy. %

{\controlpulp} requires an efficient scheme to collect sensor data at each periodic step without adding overhead to the computation part of the power management algorithm. %

\paragraph*{\textbf{{\idmae} Integration}}

As presented by Ottaviano~\etal~\cite{Ottaviano2023ControlPULPAR}, the \emph{manager domain} offloads the computation of the control action to the \emph{cluster domain}, which independently collects the relevant data from \gls{pvt} sensors and \glspl{vrm}. %
We redesign {\controlpulp}'s data movement paradigm by integrating a second dedicated {\idmae}, called \emph{\gls{sdma}}, in the \emph{manager domain} to simplify the programming model and redirect non-computational, high-latency data movement functions to the \emph{manager domain}, similar to {IBM's} \emph{Pstate} and \emph{Stop} engines~\cite{ibm_occ}. %
\rev{Our} \gls{sdma} is enhanced with \emph{rt\tus3D}, a {\me} capable of autonomously launching repeated 3D transactions. %
{\controlpulp}'s architecture is heavily inspired by {\pulpopen}\revrep{,}{;} {\idmae} integration can thus be seen in \Cref{fig:case:pulp:arch}. %
The goal of the extension is to further reduce software overhead for the data movement phase, which is beneficial to the controller's \emph{slack} within the control hyperperiod~\cite{hyperperiod}. %
The \gls{sdma} supports several interface protocols, thus allowing the same underlying hardware to handle multiple scenarios. %

\paragraph*{\textbf{Benchmarks and results}}

We evaluate the performance of the enhanced \gls{sdma} by executing the \gls{pcf} on top of FreeRTOS within an \gls{fpga}-based (Xilinx Zynq UltraScale+) \gls{hil} framework that couples the programmable logic implementing the \gls{pcs} with a power, thermal, and performance model of the plant running on top of the ARM-based Processing System~\cite{Ottaviano2023ControlPULPAR}. %

Data movement handled by \emph{rt\tus3D}, which allows repeated 3D transactions to be launched,  brings several benefits to the application scenario under analysis. %
First, it decouples the main core in the \emph{processing domain} from the \gls{sdma} in the \emph{I/O domain}. %
The \gls{sdma} autonomously realizes periodic external data accesses in hardware, minimizing the context switching and response latency suffered by the manager core in a pure software-centric approach. %
We consider a \gls{pfct} running at \SI{500}{\micro\second} and the \gls{pvct} at \SI{50}{\micro\second}, meaning at least ten task preemptions during one \gls{pfct} step with FreeRTOS preemptive scheduling policy. %
The measured task context switch time in FreeRTOS for {\controlpulp} is about 120 clock cycles~\cite{Ottaviano2023ControlPULPAR}, while \revrep{DMAE}{\idmae} programming overhead for reading and applying the computed voltages is about 100 clock cycles. %
\revrep{This saves}{From \gls{fpga} profiling runs we find that the use of \gls{sdma} saves} about 2200 execution cycles \rev{every scheduling period}\revdel{ with hardware-managed periodicity}\rev{, thus increasing the slack of the \gls{pvct} task}. %
Autonomous and intelligent data access from the \emph{I/O domain} is beneficial as it allows the two subsystems to reside in independent power and clock domains that could be put to sleep and woken up when needed, reducing the \emph{uncore} domain's power consumption. %

Our changes add minimal area overhead to the system. %
In the case of eight events and sixteen outstanding transactions, the \gls{sdma} is about \SI{11}{\kilo\gateeq} in size, accounting for an \revdel{overhead}{area increase} of only \SI{0.001}{\percent} \revrep{compared}{of} to the original {\controlpulp}\rev{'s area}. %
\revrep{This}{The} overhead \rev{imposed by \gls{sdma}} is \revrep{minimal}{negligible} when {\controlpulp} \rev{is} used as \rev{an} on-chip power manager for a large \gls{hpc} processor\rev{s.}
\rev{It} has been shown~\cite{Ottaviano2023ControlPULPAR} \rev{that the entire {\controlpulp} only} \revrep{occupy}{occupies} a \revrep{negligible}{small} area of around \SI{0.1}{\percent} on a modern \gls{hpc} CPU die. %

\subsection{{\cheshire}}
\label{sec:sys:ariane}

{\cheshire}~\cite{ottaviano2023cheshire} is a minimal, technology-independent, 64-\si{\bit} Linux-capable \gls{soc} based around {CVA6}~\cite{9220474}. %
In its default configuration, {\cheshire} features a single {CVA6} core, but coherent multicore configurations are possible. %

\paragraph*{\textbf{{\idmae} Integration}}

As {\cheshire} may be configured with multiple cores running different operating systems or {SMP} Linux, we connect our {\idmae} to the \gls{soc} using \emph{desc\tus64}. %
Descriptors are placed in \gls{spm} by a core and are then, on launch, fetched and executed by our {\idmae}. %
This \emph{single-write launch} ensures atomic operation in multi-hart environments. Support for descriptor chaining enables arbitrarily shaped transfers. %
Furthermore, transfer descriptors allow for loose coupling between the \glspl{pe} and \gls{dmae}, enabling our engine to hide the memory endpoint's latency and freeing the \glspl{pe} up to do useful work. %

\begin{figure}
    \centering%
        \subfloat{\includegraphics[width=0.95\columnwidth]{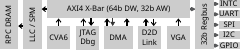}%
        \label{fig:case:cheshire:bd}%
        }

        \subfloat{\includegraphics[width=0.65\columnwidth]{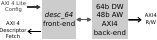}%
        \label{fig:case:cheshire:dma}%
        }
    \caption{%
        (Top) Block diagram of the {\cheshire} \gls{soc}. %
        (Bottom) Configuration system's \gls{axi} \revrep{DMA}{\idmae}. %
    }%
    \label{fig:case:cheshire:arch}%
\end{figure}

The used {\be} is configured to a data and address width of \SI{64}{\bit} and can track eight outstanding transactions, enough to support efficient fine-grained accesses to the \gls{spm} and external memory IPs. %
A schematic view of {\cheshire} and the {\idmae} configuration can be found in \Cref{fig:case:cheshire:arch}. %

\paragraph*{\textbf{Benchmarks}}

We use the {AXI DMA v7.1}~\cite{amd_xilinx_axi_2022} from Xilinx integrated into the {\cheshire} \gls{soc} as a comparison. %
We run synthetic workloads copying data elements of varying lengths, allowing a more direct comparison of the bus utilization at a given transfer length. %

\paragraph*{\textbf{Results}}

\begin{figure}
    \centering%
    \includegraphics[width=\columnwidth]{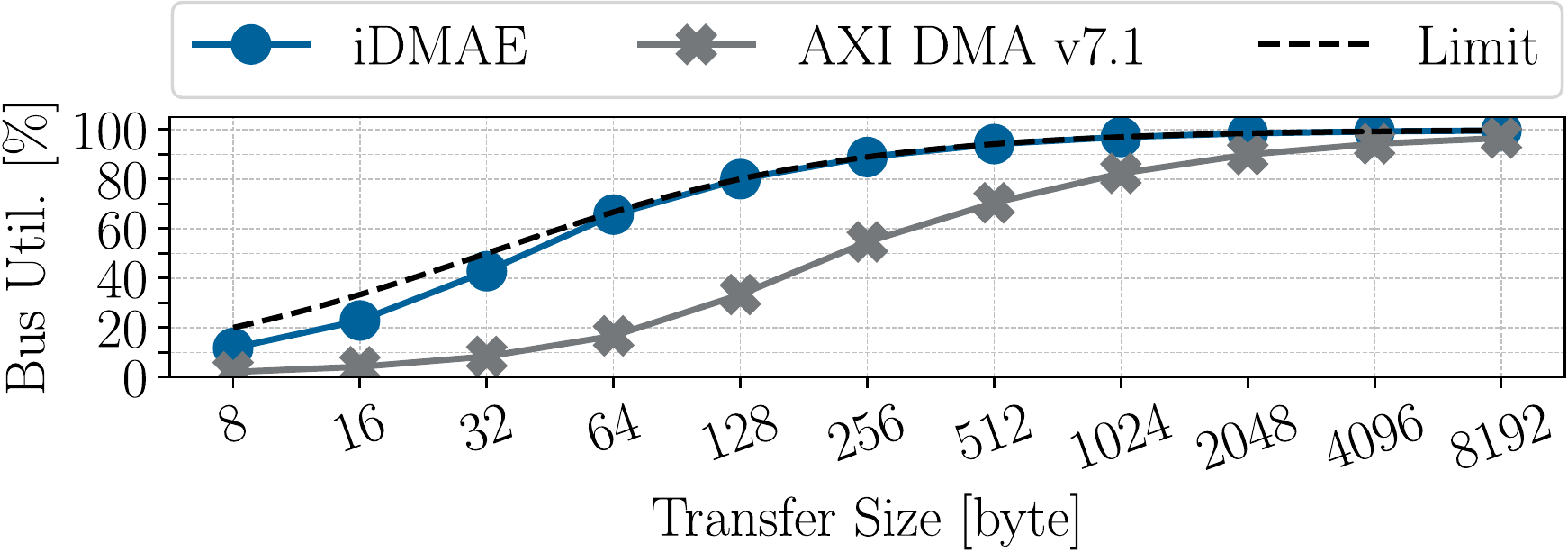}%
    \caption{%
        Bus utilization given a certain length of transfers. %
        The performance of our {\idmae} is compared to Xilinx's {AXI DMA v7.1}~\cite{amd_xilinx_axi_2022}. %
        The dotted line represents the theoretical limit physically possible. %
    }%
    \label{fig:case:ariane-util}%
\end{figure}

Compared to {AXI DMA v7.1}, {\idmae} increases bus utilization by %
almost 6$\times$ when launching fine-grained 64~\si{\byte} transfers. %
At this granularity, {\idmae} achieves almost perfect utilization, as shown in \Cref{fig:case:ariane-util}. %
We implemented both designs on a {Diligent} {Genesis II} \glsunset{fpga}\gls{fpga}, reducing the required {LUTs} by over \SI{10}{\percent} and {FFs} by over \SI{23}{\percent}. %
As our {\idmae} does not require \gls{sram} buffers, we can reduce the amount of {BRAMs} used from \SI{216}{\kibi\bit} to zero. %

\subsection{\mempool{}}
\label{sec:sys:mempool}

\mempool{}~\cite{Riedel2023} is a flexible and scalable single-cluster manycore architecture featuring 256 32-\si{\bit} RISC-V cores that share \SI{1}{\mebi\byte} of low-latency \gls{l1} \gls{spm} distributed over 1024 banks. %
All cores are individually programmable, making \mempool{} well-suited for massively parallel regular workloads like computational photography or machine learning and irregular workloads like graph processing. %
The large shared \gls{l1} memory simplifies the programming model as all cores can directly communicate via shared memory without explicit dataflow management. %
The \gls{l1} banks are connected to the cores via a pipelined, hierarchical interconnect. %
Cores can access banks close to them within a single cycle, while banks further away have a latency of three or five cycles. %
In addition to the \gls{l1} interconnect, the cores have access to a hierarchical \glsunset{axi}\gls{axi}~\cite{arm_amba_2021} interconnect connecting to the \gls{soc}. %

\paragraph*{\textbf{{\idmae} Integration}}

\mempool{}'s large scale and distributed \gls{l1} memory make a monolithic \dmae{} incredibly expensive as it would require a dedicated interconnect, spanning the whole \mempool{} architecture, connecting all 1024 memory banks. %
The existing interconnect between cores and \gls{l1} memory is built for narrow, single-word accesses; thus unsuitable for wide, burst-based transfers. %

To minimize interconnect overhead to \gls{l1} memory, multiple {\bes} are introduced into \mempool{} placed close to a group of banks. %
To connect to the \gls{soc}, it can share the existing \gls{axi} interconnect used to fetch instructions. %

These distributed \idmae{} {\bes}, each controlling exclusive regions of the \gls{l1} memory, greatly facilitate physical implementation. %
However, individually controlling all \bes{} would burden the programmer and massively increase overhead due to transfer synchronization of the individual {\dmaes}. %
Instead, our \idmae{}'s modular design allows for hiding this complexity in hardware by using a single \fe{} to program all the distributed \bes{}. %

\begin{figure}
    \centering%
    \includegraphics[width=\columnwidth]{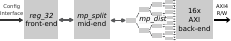}%
    \caption{%
        Our distributed {\idmae} implemented in \mempool{}. %
        \emph{Mp{\tus}split} splits the transfers along their \gls{l1} boundaries, and a tree of \emph{mp{\tus}dist} \mes{} distribute the transfer. %
    }%
    \label{fig:case:mempool}%
\end{figure}

As seen in \Cref{fig:case:mempool}, the \emph{mp{\tus}split} {\me} splits a single \dma{} request along lines of \mempool{}'s \gls{l1} memory's address boundaries, and a binary tree of \emph{mp{\tus}dist} {\mes} distributes the resulting requests to all \bes{}. %

\paragraph*{\textbf{Benchmarks}}

We evaluate \mempool{}'s \idmae{} by comparing the performance of various kernels compared to a baseline without {\dma}. %
Since \mempool{} requires our modular \idmae{} to implement a distributed \dma{}, a comparison with another \dma{} unit is not feasible here. %
First, we compare the performance copying \SI{512}{\kibi\byte} from \gls{l2} to \gls{l1} memory. %
Without a \dma{}, the cores can only utilize \revrep{a}{one} sixteenth of the wide \gls{axi} interconnect. %
The \idmae{} utilizes \SI{99}{\percent} and speeds up memory transfers by a factor of 15.8\x~while incurring an area overhead of less than \SI{1}{\percent}. %

The performance improvement for kernels is evaluated by comparing a double-buffered implementation supported by our \idmae{} to the cores copying data in and out before and after the computation. %
Even for heavily compute-bound kernels like matrix multiplication, {\idmae} provides a speedup of 1.4\x. %
Less compute-intensive kernels like the convolution or discrete cosine transformation benefit even more from the \idmae{} with speedups of 9.5\x~and 7.2\x, respectively. %
Finally, memory-bound kernels like vector addition and the dot product are dominated by the data transfers and reach speedups of 15.7\x~and 15.8\x. %

\subsection{{\occamy}}
\label{sec:sys:occamy}

{\occamy}{\occamyfn} is a high-performance compute platform based on the Manticore~\cite{9220474} 2.5D chiplet concept. %
It provides two Linux-capable {CVA6}~\cite{9220474} host cores and 432 Snitch~\cite{Zaruba_2021} worker cores grouped in 48 compute clusters and sharing \SI{16}{\gibi\byte} of \gls{hbm} across two dies. %
It enhances its RISC-V \gls{isa} with lightweight extensions to maximize its \gls{fpu} utilization in both regular and irregular workloads. 
{\occamy}'s use cases range from DNN training and inference to stencil codes and sparse scientific computing. %

\paragraph*{\textbf{{\idmae} Integration}}

Each Snitch cluster has an {\idmae}, \rev{called} \emph{cluster {\dma}}, fetching data directly from \gls{hbm} or sharing data between clusters; a complex hierarchical interconnect is required to support both efficiently. %
The cluster {\dma} present in the \gls{soc} is used only to manage operations and initialize memory. %

\begin{figure}
    \centering%
        \subfloat{\includegraphics[width=0.95\columnwidth]{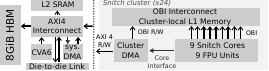}%
        \label{fig:case:occamy:bd}%
        }

        \subfloat{\includegraphics[width=0.8\columnwidth]{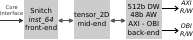}%
        \label{fig:case:occamy:dma}%
        }
    \caption{%
        (Top) Block diagram of one {\occamy} die. %
        (Bottom) Configuration of the cluster {\dma}. %
    }%
    \label{fig:case:occamy:snitch}%
\end{figure}

Each {\idmae} is tightly coupled to a \emph{data movement} core, as shown in \Cref{fig:case:occamy:snitch}. %
The Snitch core decodes the \revrep{DMA}{\idma} instructions and passes them to an \emph{inst\tus64} {\fe}. %
A \emph{tensor{\tus}ND} {\me} enables efficient 2D affine transfers. %
Higher-dimensional or irregular transfers can be handled efficiently through fine-granular control code on the core: configuring and launching a 1D transfer incurs only three instructions, while 2D transfers require at most six instructions to be launched. %

The \emph{cluster {\dma}} is configured with a data width of \SI{512}{\bit} and an address width of \SI{48}{\bit}. %
It can track 32 outstanding transactions, enabling efficient transfer and latency hiding on fine-granular accesses to the long-latency \gls{hbm} endpoint, even in the face of congestion from other clusters. %
It provides one AXI4 read-write port connecting to the surrounding \gls{soc} and one OBI read-write port connecting to the cluster's banked \gls{l1} memory. %
\gls{axi} intraprotocol transfers are enabled to reorganize data \revrep{from}{within} HBM or \gls{l1}. %

\paragraph*{\textbf{Benchmarks}}

We evaluate \gls{gemm}, \gls{spmv}, and \gls{spmm} on {\occamy} with and without the use of cluster DMA engines. %
We run \gls{rtl} simulation on clusters processing double-precision tiles and use these results to compute the performance of a single chiplet, taking into account bandwidth bottlenecks and assuming all reused data is ideally cached. %

Each workload is evaluated with four cluster tile sizes \emph{S}, \emph{M}, \emph{L}, and \emph{XL}; for \gls{gemm}, these are square tiles of size 24, 32, 48, and 64, while the two sparse workloads use the matrices of increasing density \emph{diag}, \emph{cz2548}, \emph{bcsstk13}, and \emph{raefsky1} from the SuiteSparse matrix collection~\cite{Davis2011TheUO} as tiles. %

As Snitch~\cite{Zaruba_2021} originally does not include a {\dmae}, we compare {\occamy} to an architecture where worker cores make all data requests with ideal capability to handle outstanding transactions, but real bandwidth limitations. %

\paragraph*{\textbf{Results}}

\begin{figure}
    \centering%
    \includegraphics[width=\columnwidth]{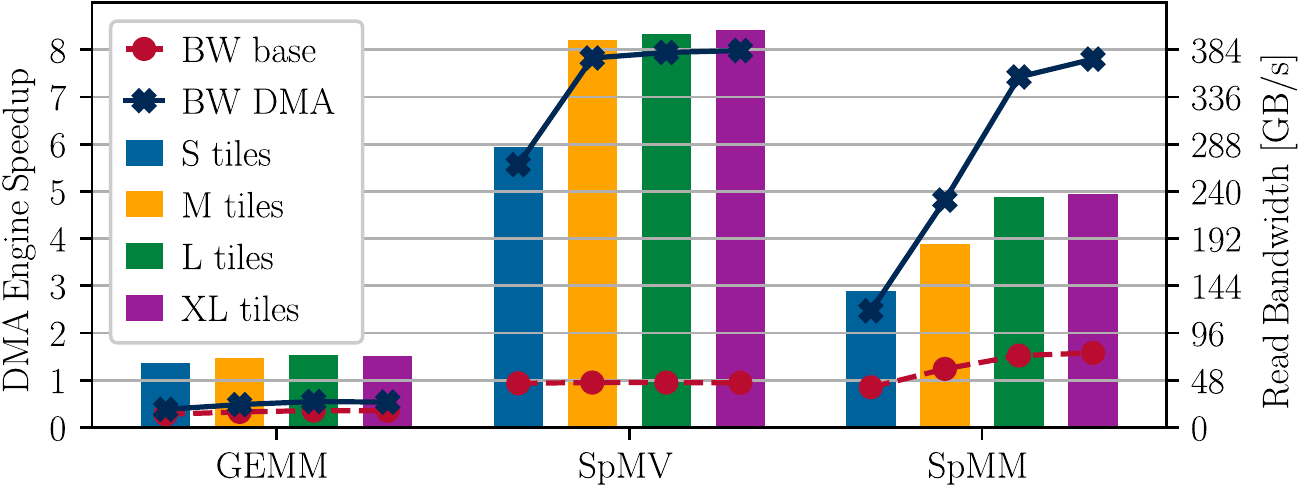}%
    \caption{%
        {\occamy} chiplet bandwidths and speedups enabled by {\idma} on workloads with varying tile sizes. %
    }%
    \label{fig:case:occamy:results}%
\end{figure}

\Cref{fig:case:occamy:results} shows our results. %
For the highly compute-bound \gls{gemm}, our \revrep{DMAEs}{\idmaes} enable moderate, but significant speedups of 1.37\x~to 1.52\x. %
Since all tile sizes enable ample cluster-internal data reuse, we see only small benefits as tiles grow. %
Nevertheless, the cluster engines still increase peak HBM read bandwidth from 17 to \SI{26}{\giga\byte\per\second}. %

\gls{spmv} performance, on the other hand, is notoriously data-dependent and memory-bound due to \rev{a} lack of data reuse. %
Unable to leverage on-chip caches fed by a wider network, the baseline nearly saturates its narrow interconnect for all tile sizes at \SI{48}{\giga\byte\per\second}. %
The {\idmaes} only become memory bound past \emph{M}-sized tiles, but then approach the wide interconnect peak throughput of \SI{384}{\giga\byte\per\second}. %
Overall, the engines enable significant speedups of 5.9\x~to 8.4\x. %

\gls{spmm} is similar to \gls{spmv}, but enables on-chip matrix data reuse, becoming compute-bound for both the baseline and {\idmaes}. %
Since data caching is now beneficial, the baseline overcomes the \SI{48}{\giga\byte\per\second} bottleneck and speedups decrease \revrep{to}{from} 2.9\x~to 4.9\x. %
Still, {\idmaes} unlock the full compute of clusters on sparse workloads while approaching the  \SI{384}{\giga\byte\per\second} peak throughput only for \emph{XL} tiles. %

\subsection{\rev{{\idma} Integration and Customization}}
\label{sec:integration}
\rev{%
The high degree of parameterization of {\idma} might impose a steep learning curve on any new designer implementing {\idma} into a system. %
To ease the process, we provide \emph{wrapper modules}, which abstract internal, less-critical parameters away and only expose a critical selection of parameters: \emph{address width (AW)}, \emph{data width (DW)}, and \emph{number of outstanding transactions (NAx)} to the user. %
\Cref{sec:results} discusses the influence of these parameters on area, timing, latency, and performance. %
}

\rev{%
Customizing {\idma} to a given target system involves two main steps: %
analyzing the system's transfer pattern and either selecting the required {\me}(s) or creating a new one and selecting an appropriate parametrization. %
The \emph{(NAx)} parameter should be selected high enough to saturate the memory system when launching the finest-granular transfers while not overwhelming the downstream targets. %
}

\section{Architecture Results}
\label{sec:results}

\newcommand{\rott}[1]{\rotatebox[origin=c]{90}{#1}}
\newcommand{\divl}{\noalign{\smallskip} \arrayrulecolor{gray}\cline{2-15} \noalign{\smallskip}}

\newcommand{\ctl}[2]{\makecell[cl]{#1 \\ #2}}

\newcommand{\res}[4]{\makecell[cl]{\SI{#1}{#3}{#4} \\ $\mathcal{O}$\textit{(#2)}}}
\newcommand{\resc}[4]{\SI{#1}, $\mathcal{O}$\textit{(#2)}{#4}}
\newcommand{\resh}[4]{\SI{#1}{#3}{#4}}
\newcommand{\reshc}[4]{\SI{#1}{#4}}
\newcommand{\resn}{\textcolor{gray}{\textit{0}}}

\begin{table*}
    \caption{%
    Area decomposition of the {\dmae} configuration used in the \gls{pulp}-cluster, see \cref{sec:sys:pulp}. %
    The \emph{base} area is always required, the contribution of each protocol added is shown. %
    If the area contribution is non-zero, the parameter influencing the value is provided using the \emph{big-$\mathcal{O}$} notation. %
    The area contribution scales linearly with the data width (\textit{DW}) if no scaling is provided. %
    }
    \label{tab:results:area}
    \renewcommand*{\arraystretch}{2.2}
    \resizebox{2.05\columnwidth}{!}{%
    \begin{threeparttable}
        \begin{tabular}{@{}l!{\color{black}\vrule}l!{\color{black}\vrule}llllllllllllll@{}} \toprule
            \multicolumn{3}{l}{Units} &
            Base &
            \multicolumn{2}{c}{\makecell[cc]{AXI \\ read \hspace{0.5cm} write}} &
            \multicolumn{2}{c}{\makecell[cc]{AXI Lite\\ read \hspace{0.5cm} write}} &
            \multicolumn{2}{c}{\makecell[cc]{AXI Stream\\ read \hspace{0.5cm} write}} &
            \multicolumn{2}{c}{\makecell[cc]{OBI \\ read \hspace{0.5cm} write}} &
            \multicolumn{2}{c}{\makecell[cc]{TileLink \\ read \hspace{0.5cm} write}} &
            Init \\
            \midrule
            \multirow{8}{*}{\rott{Backend}} &
            \rott{Decoupling} &
            - &
            \res{3.7}{NAx}{\kilo\gateeq}{~\tnote{a}} &
            \res{1.4}{NAx}{\kilo\gateeq}{} &
            \res{1.4}{NAx}{\kilo\gateeq}{} &
            \res{310}{NAx}{\gateeq}{} &
            \res{310}{NAx}{\gateeq}{} &
            \res{310}{NAx}{\gateeq}{} &
            \res{310}{NAx}{\gateeq}{} &
            \res{310}{NAx}{\gateeq}{} &
            \res{310}{NAx}{\gateeq}{} &
            \res{310}{NAx}{\gateeq}{} &
            \res{310}{NAx}{\gateeq}{} &
            \resn \\ \divl
            &
            \multirow{4}{*}{\rott{Legalizer}} &
            State &
            \res{1.5}{AW}{\kilo\gateeq}{~\tnote{b}} &
            \res{710}{AW}{\gateeq}{~\tnote{c}} &
            \res{710}{AW}{\gateeq}{~\tnote{c}} &
            \res{200}{AW}{\gateeq}{~\tnote{c}} &
            \res{200}{AW}{\gateeq}{~\tnote{c}} &
            \res{180}{AW}{\gateeq}{~\tnote{c}} &
            \res{180}{AW}{\gateeq}{~\tnote{c}} &
            \res{180}{AW}{\gateeq}{~\tnote{c}} &
            \res{180}{AW}{\gateeq}{~\tnote{c}} &
            \res{215}{AW}{\gateeq}{~\tnote{c}} &
            \res{215}{AW}{\gateeq}{~\tnote{c}} &
            \res{21}{AW}{\gateeq}{} \\
            &
            &
            Page Split &
            \resn &
            \res{95}{1}{\gateeq}{} &
            \res{105}{1}{\gateeq}{} &
            \res{7}{1}{\gateeq}{} &
            \res{8}{1}{\gateeq}{} &
            \resn &
            \resn &
            \res{5}{1}{\gateeq}{} &
            \res{5}{1}{\gateeq}{} &
            \resn &
            \resn &
            \resn \\
            &
            &
            Pow2 Split &
            \resn &
            \resn &
            \resn &
            \resn &
            \resn &
            \resn &
            \resn &
            \resn &
            \resn &
            \res{20}{1}{\gateeq}{} &
            \res{20}{1}{\gateeq}{} &
            \resn \\ \divl
            &
            \multirow{4}{*}{\rott{Transport Layer}} &
            Dataflow Element &
            \resh{1.3}{DW}{\kilo\gateeq}{~\tnote{d}} &
            \resn &
            \resn &
            \resn &
            \resn &
            \resn &
            \resn &
            \resn &
            \resn &
            \resn &
            \resn &
            \resn \\
            &
            &
            \makecell[cl]{Contribution \\ of \revdel{e}{E}ach Read/ \\ Write Manager \\ Respectively}&
            \resh{70}{DW}{\gateeq}{} &
            \resh{190}{DW}{\gateeq}{} &
            \resh{30}{DW}{\gateeq}{} &
            \resh{60}{DW}{\gateeq}{} &
            \resh{60}{DW}{\gateeq}{} &
            \resh{60}{DW}{\gateeq}{} &
            \resh{60}{DW}{\gateeq}{} &
            \resh{60}{DW}{\gateeq}{} &
            \resh{35}{DW}{\gateeq}{} &
            \resh{230}{DW}{\gateeq}{} &
            \resh{150}{DW}{\gateeq}{} &
            \resh{55}{DW}{\gateeq}{} \\
            &
            &
            Shifter/Muxing &
            \resh{120}{DW}{\gateeq}{} &
            \resh{250}{DW}{\gateeq}{~\tnote{c}} &
            \resh{250}{DW}{\gateeq}{~\tnote{c}} &
            \resh{75}{DW}{\gateeq}{~\tnote{c}} &
            \resh{75}{DW}{\gateeq}{~\tnote{c}} &
            \resh{180}{DW}{\gateeq}{~\tnote{c}} &
            \resh{180}{DW}{\gateeq}{~\tnote{c}} &
            \resh{170}{DW}{\gateeq}{~\tnote{c}} &
            \resh{170}{DW}{\gateeq}{~\tnote{c}} &
            \resh{65}{DW}{\gateeq}{~\tnote{c}} &
            \resh{65}{DW}{\gateeq}{~\tnote{c}} &
            \resn \\
            \bottomrule
        \end{tabular}
        \begin{tablenotes}[para, flushleft]
            \item[a] \emph{NAx}: Number of outstanding transfers supported. Used configuration: \textbf{16}.
            \item[b] \emph{AW}: Address width. Used configuration: \textbf{32-\si{\bit}}.
            \item[c] If multiple protocols are used, only the maximum is taken.
            \item[d] \emph{DW}: Data Width. Used configuration: \textbf{32-\si{\bit}}.
        \end{tablenotes}
    \end{threeparttable}
    }
\end{table*}

To deepen the insight into {\idma} and highlight its versatility, we provide \glsu{ip}-level implementation results in this section. %
We first present area and timing models characterizing the influence of parametrization on our architecture, enabling quick and accurate estimations when integrating engines into new systems. %
We then use these models to show that {\idma}'s area and timing scale well for any reasonable parameterization. %
Finally, we present latency results for our {\be} and discuss our engine's performance in three sample memory systems. %
For implementation experiments, we use {\gfs} {\emph{\gftech}} technology with a 13-metal stack and 7.5-track standard cell library in the typical process corner. %
We synthesize our designs using {\dc} in topological mode to account for place-and-route constraints, congestion, and physical phenomena. %

\begin{figure}
     \centering
        \subfloat{\includegraphics[width=\columnwidth]{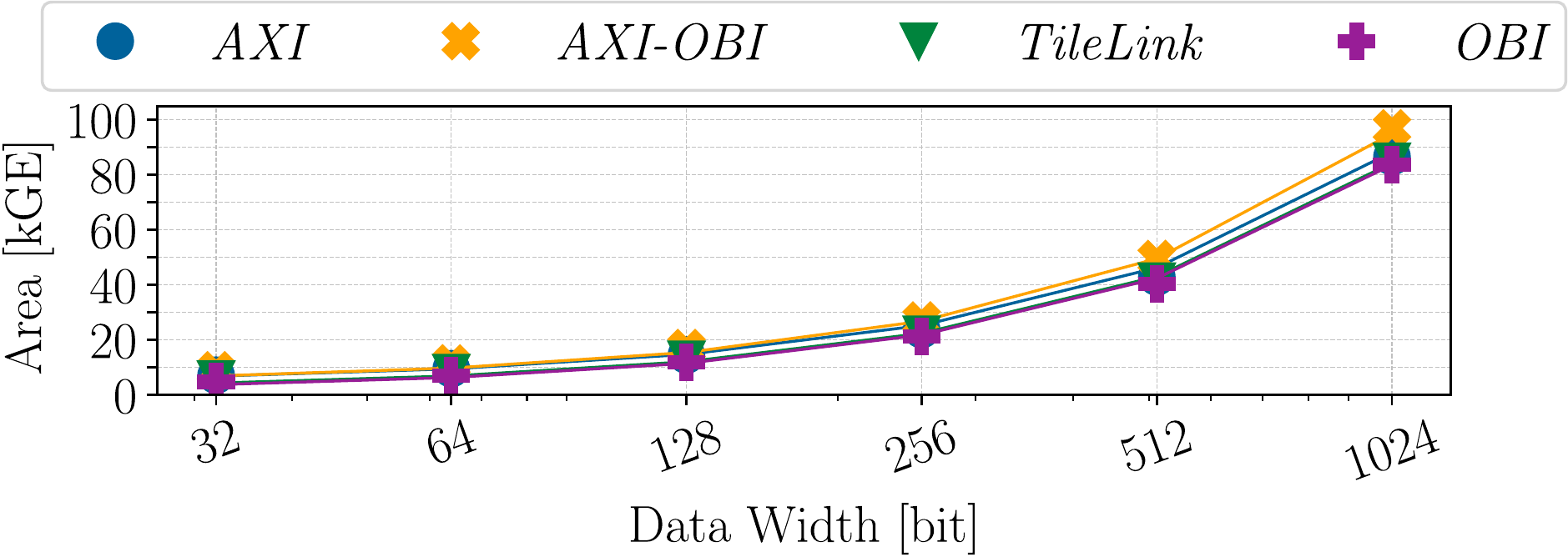}%
        \label{fig:results:area:data-width}%
        }

        \subfloat{\includegraphics[width=0.48\columnwidth]{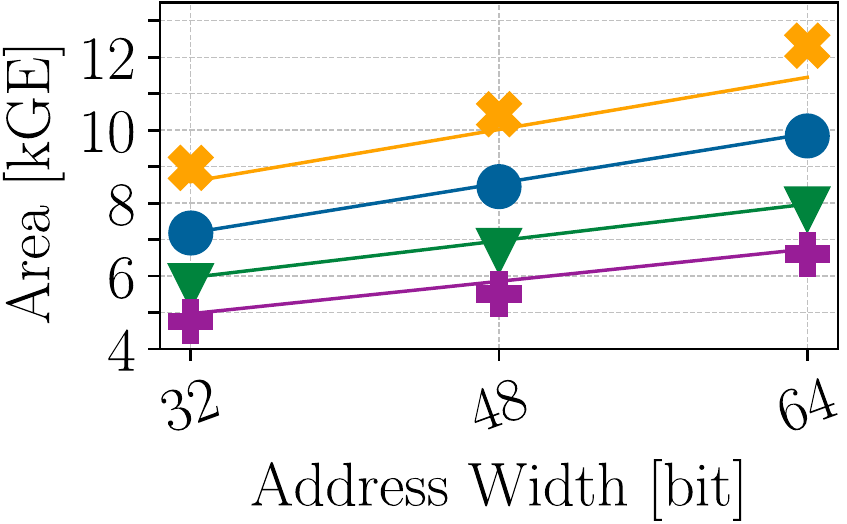}%
        \label{fig:results:area:addr-width}%
        }
        \subfloat{\includegraphics[width=0.48\columnwidth]{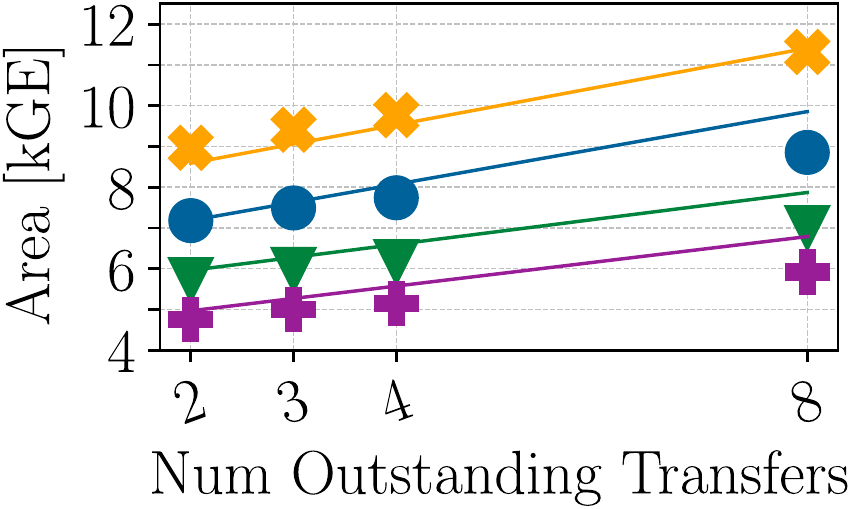}%
        \label{fig:results:area:num-ax}%
        }
     \caption{ %
        Area scaling of a {\be} base configuration (\emph{32-\si{\bit}} address and data width, \emph{two} outstanding transactions). %
         Markers represent the measurement points and lines the fitted model. %
     } %
     \label{fig:results:area}
\end{figure}

\subsection{Area Model}
\label{sec:results:area}

We focus our evaluation and modeling effort on our {\idma} {\be}, as both the {\fe} and {\me} are very application- and platform-specific and can only be properly evaluated in-system. %
Our area model was evaluated at \SI{1}{\giga\hertz} using the typical process corner in {\gftech}. %

For each of the {\be}'s major area contributors listed in \Cref{tab:results:area}, we fit a set \rev{of} linear model\rev{s} using \emph{non-negative least squares}. %
For each parametrization, our models take a vector containing the number of ports of each protocol as an input. %
This set of models allows us to estimate the area decomposition of the base hardware of the {\be} and the area contributions of any additional protocol port, given a particular parameterization, with an average error of less than \SI{4}{\percent}. %
For example, \Cref{tab:results:area} shows the modeled area decomposition for our \emph{base} configuration of 32-\si{\bit} address width, \rev{32-\si{\bit}} data width, and two outstanding transfers. %

A second step is required to estimate area contributions to the {\be} depending on the parameterization, the number, and \rev{the} type of ports. %
We created a second \emph{param} model estimating the influence of \rev{the} three main parameters, \emph{area width (AW)}, \emph{data width (DW)}, and the \emph{number of outstanding transactions (NAx)}, on the {\be}'s area contributions. %
We can estimate the area composition of the {\be} with an average error of less than \SI{9}{\percent}, given both the parameterization and the used read/write protocol ports as input. %

We provide a qualitative understanding of the influence of parameterization on area by listing the parameter with the strongest correlation using \emph{big-$\mathcal{O}$} notation in \Cref{tab:results:area}. %

To outline the accuracy of our modeling approach, we show the area scaling of four of our {\idmaes} for different protocol configurations, depending on the three main parameters, starting from the \emph{base} configuration. %
The subplots of \Cref{fig:results:area} present the change in area when one of the three main parameters is modified and the output of our two linear models combined. %
The combined area model tracks the parameter-dependent area development with an average error of less than \SI{9}{\percent}. %
In those cases where the model deviates, the modeled area is overestimated, providing a safe upper bound for the {\be} area. %

\subsection{Timing Model}
\label{sec:results:timing}

\begin{figure}
     \centering
        \subfloat{\includegraphics[width=\columnwidth]{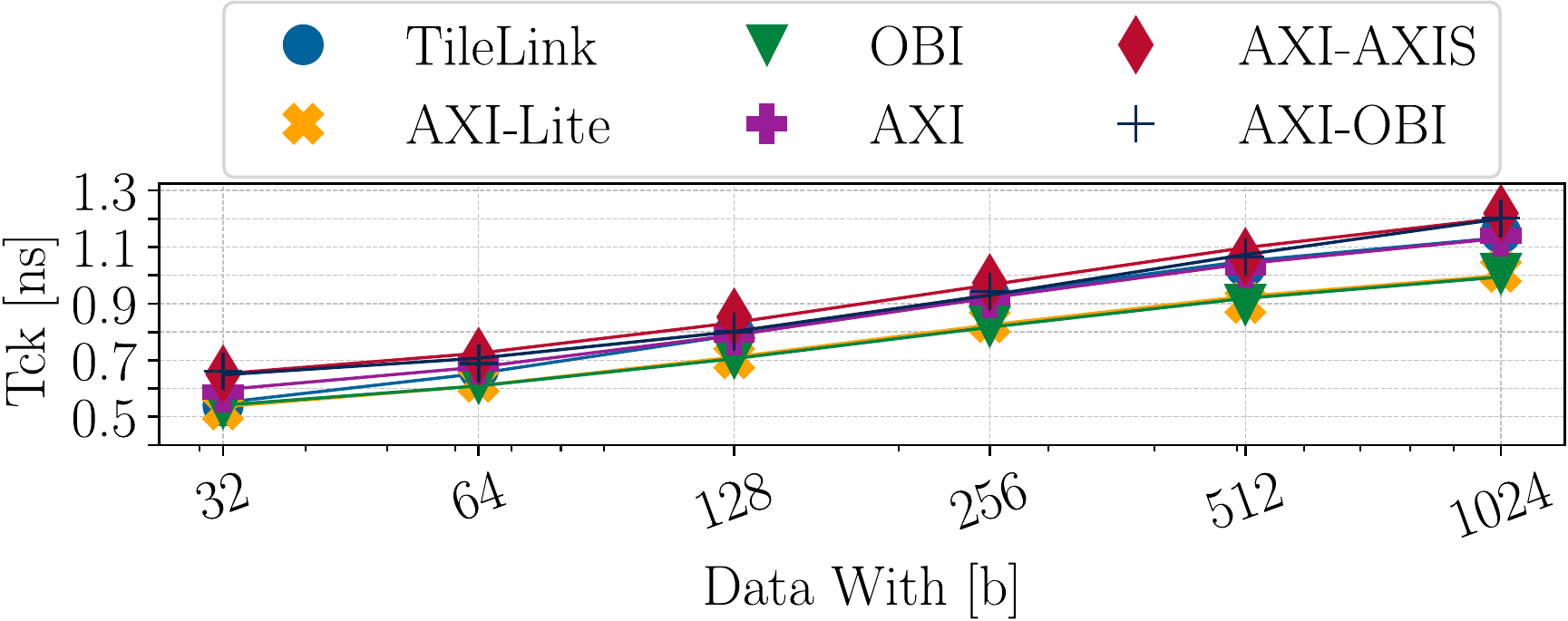}%
        \label{fig:results:tck:data-width}%
        }

        \subfloat{\includegraphics[width=0.48\columnwidth]{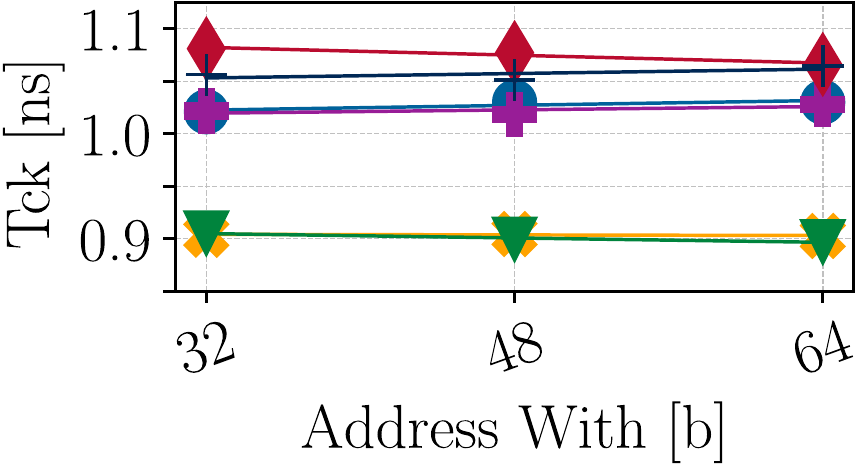}%
        \label{fig:results:tck:addr-width}%
        }
        \subfloat{\includegraphics[width=0.48\columnwidth]{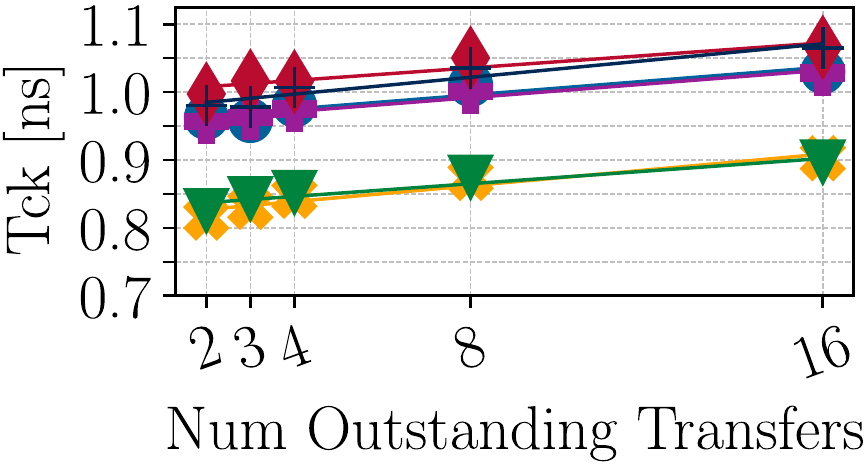}%
        \label{fig:results:tck:num-ax}%
        }
     \caption{Clock frequency scaling of a {\be} base configuration (\emph{32-\si{\bit}} address and data width, \emph{two} outstanding transactions).}
     \label{fig:results:tck}
\end{figure}

We again focus our timing analysis on the {\be}, as the {\fe} should be analyzed in-system and {\mes} may be isolated from the {\idmae}'s timing by cutting timing paths between \revrep{the mid-end and back-end}{front-, mid-, and back-ends}. %
Our investigation shows a \emph{multiplicative inverse} dependency between the longest path in \emph{\si{\nano\second}} and our main parameters. %
We use the \emph{base} configuration of the {\be} to evaluate our timing model by sweeping our three main parameters. %
The tracking of our model is presented in \Cref{fig:results:tck} for six representative configurations ranging from simple \gls{obi} to complex multi-protocol configurations involving \gls{axi}. %
Our timing model achieves an average error of less than \SI{4}{\percent}. %

The results divide our {\bes} into two groups: simpler protocols, \emph{OBI} and \emph{AXI Lite}, run faster as they require less complex legalization logic, whereas more complex protocols require deeper logic and thus run slower. %
Engines supporting multiple protocols and ports also run slower due to additional arbitration logic in their data path. %
\emph{Data width} has a powerful impact on {\idmae}'s speed, mainly due to wider shifters required to align the data. %
The additional slowdown at larger data widths can be explained by physical routing and placement congestion of the increasingly large buffer in the dataflow element. %
\emph{Address width} has little effect on the critical path as it does not pass through the legalizer cores, whose timing is most notably affected by address width. %
Increasing the number of \emph{outstanding transactions} sub-linearly degrades timing due to more complex \gls{fifo} management logic required to orchestrate them. %

\subsection{Latency}
\label{sec:results:latency}

Our {\idma} {\bes} have a fixed latency of \emph{two} cycles from receiving a 1D transfer from the {\fe} or the last {\me} to the read request at a protocol port. %
Notably, this is \emph{independent} of the protocol selection, the number of protocol ports, and the three main {\idma} parameters. %
This rule only has one exception: in a {\be} without hardware legalization support, the latency reduces to \emph{one} cycle. %
Generally, each {\me} presented in \Cref{tab:arch:me} requires \emph{one} additional cycle of latency. We note, however, that the \emph{tensor{\tus}ND} {\me} can be configured to have \emph{zero} cycles of latency, meaning that even for an N-dimensional transfer, we can ensure that the first read request is issued \emph{two} cycles after the transfer arrives at the {\me} from the {\fe}. %

\subsection{Standalone Performance}
\label{sec:results:perf}

\begin{figure}
     \centering
        \subfloat{\includegraphics[width=0.85\columnwidth]{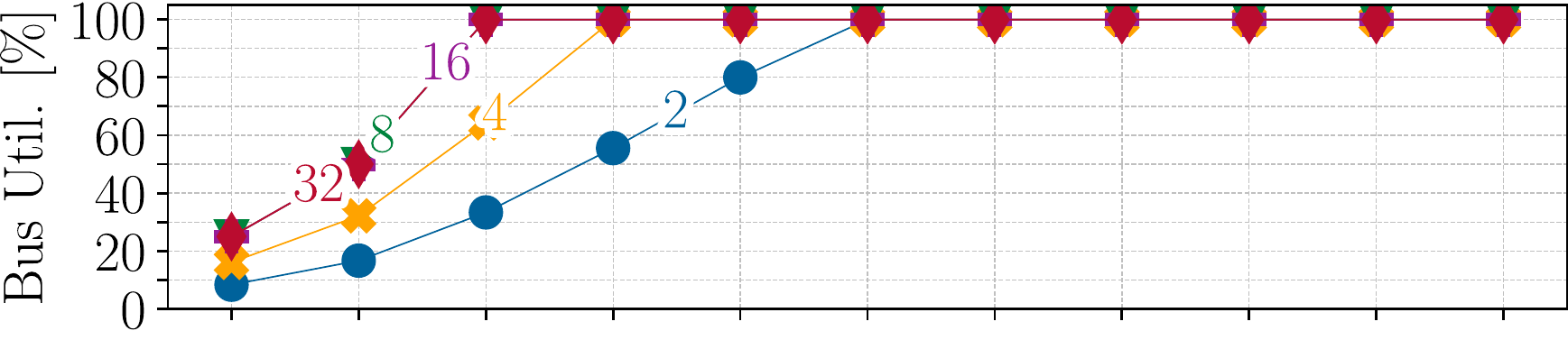}%
        \label{fig:results:perf:pulp}%
        }

        \subfloat{\includegraphics[width=0.85\columnwidth]{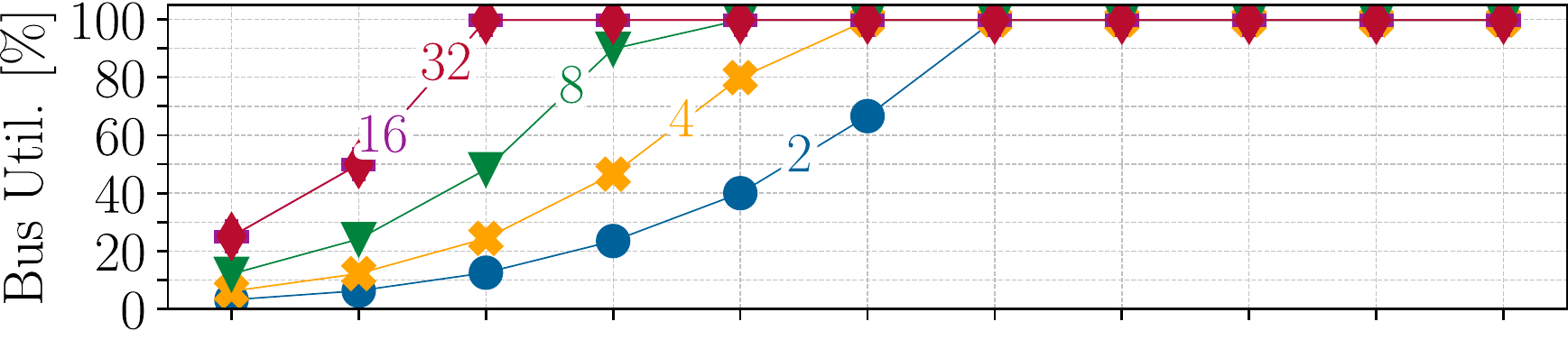}%
        \label{fig:results:perf:rpc}%
        }

        \subfloat{\includegraphics[width=0.85\columnwidth]{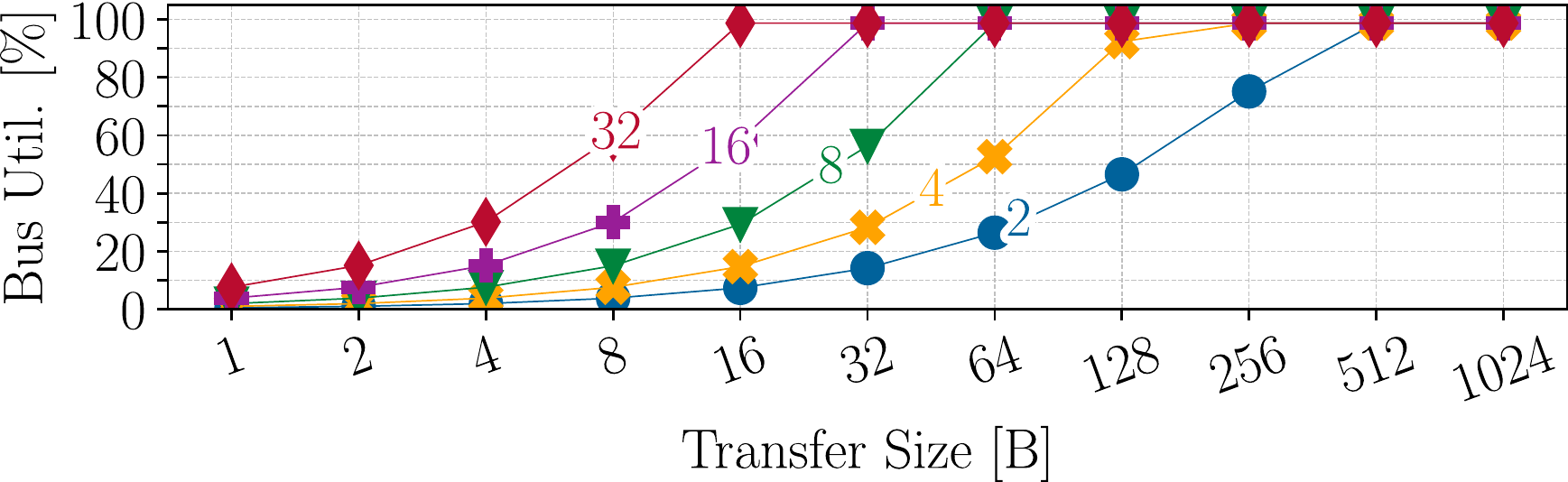}%
        \label{fig:results:perf:hbm}%
        }
     \caption{%
         Bus utilization of our {\idmae} in the base configuration (\emph{32-\si{\bit}} address and data width) with varying amounts of outstanding transactions in three different memory systems; \emph{SRAM}, \emph{\gls{rpc} DRAM}~\cite{rpc-dram-datasheet}, \emph{HBM}~\cite{9220474}. %
      }
     \label{fig:results:perf}
\end{figure}

We evaluated the out-of-context performance of an {\idmae} in the \emph{base} configuration copying a \emph{\SI{64}{\kibi\byte}} transfer fragmented in individual transfer sizes between \SI{1}{\byte} and \SI{1}{\kibi\byte} in three different memory system models. %
The analysis is protocol-agnostic as all implemented protocols support a similar outstanding transaction mechanism; we thus use \emph{AXI4} in this subsection without loss of generality. %
The three memory systems used in our evaluation differ in access cycle latency and number of outstanding transfers. %

\emph{SRAM} represents the \gls{l2} memory found in the PULP-open system (\Cref{sec:sys:pulp}) with three cycles of latency and eight outstanding transfers. %
\emph{\gls{rpc}-DRAM} uses the characteristics of an open-source AXI4 controller for the \gls{rpc} DRAM technology~\cite{rpc-dram-flyer,rpc-dram-datasheet} run at \SI{933}{\mega\hertz}, around thirteen cycles of latency and support for sixteen outstanding transactions. %
\emph{\gls{hbm}} models an industry-grade \gls{hbm}~\cite{9220474} interface with a latency in the order of 100 cycles and supporting the tracking of more than 64 outstanding transfers. %

In shallow memory systems, the {\idmae} reaches almost perfect bus utilization copying single bus-sized data transfers required to track as low as eight outstanding transactions. %
More outstanding requests are required in deeper memory systems to sustain this perfect utilization. %
Any transfers smaller than the bus width will inevitably lead to a \rev{\emph{substantial}} drop in utilization, meaning that unaligned transfers inherently limit the maximum possible bus utilization our engines can achieve. Nevertheless, our fully decoupled data-flow-oriented architecture \rev{\emph{maximizes the utilization of the bus}} even in these scenarios. %

\Cref{fig:results:perf} shows that even in very deep systems with hundreds of cycles of latency, our engine can achieve almost perfect utilization for a relatively small transfer granularity of four times the bus width. %
This agility in handling transfers allows us to copy multi-dimensional tensors with a narrow inner dimension efficiently. %

The cost of supporting such fine-granular transfers is an increased architectural size of the engines' decoupling buffers. %
As shown in \Cref{fig:results:area:num-ax}, these scale linearly in the number of outstanding transactions to be supported, growing by roughly \SI{400}{\gateeq} for each added buffer stage. %
In our base configuration, supporting 32 outstanding transfers keeps the engine area below \SI{25}{\kilo\gateeq}. %

\subsection{\rev{Energy Efficiency}}
\label{sec:results:energy}

\rev{%
With {\idma}'s area-optimized design, the capability of selecting the simplest on-chip protocol, and the effort to minimize buffer area, we can curtail the area footprint and thus the static power consumption of our {\idmaes}. %
}

\rev{%
{\idma}'s decoupled, agile, data-flow-oriented architecture is explicitly designed to handle transfers efficiently while maximizing bus utilization, limiting the unit's active time to a minimum. %
Coupled with our minimal area footprint and low-buffer design, this directly minimizes energy consumption. %
Furthermore, the engine's efficiency allows \emph{run-to-completion} operating modes where we maximize the interconnect's periods of inactivity between transfers, allowing efficient clock gating of the {\idmae} and the interconnect, further increasing energy efficiency. %
}

\section{Related Work}
\label{sec:related-work}

\newcommand{\hol}[1]{\makecell[cc]{\textbf{#1}}}
\newcommand{\htl}[2]{\makecell[cc]{\textbf{#1} \\ \textbf{#2}}}
\newcommand{\hhl}[3]{\makecell[cc]{\textbf{#1} \\ \textbf{#2} \\ \textbf{#3}}}

\newcommand{\did}[2]{\vspace{0.05cm}\makecell[cc]{{#1} \\ \emph{#2}}\vspace{0.05cm}}
\newcommand{\dis}[1]{\vspace{0.05cm}\makecell[cc]{\emph{#1}}\vspace{0.05cm}}
\newcommand{\dtw}[2]{\vspace{0.05cm}\makecell[cc]{\textbf{#1} \\ \emph{#2}}\vspace{0.05cm}}

\begin{table*}[p]
    \scriptsize{%
        \centering
        \caption{%
            Comparison of \revrep{our DMA architecture}{\idma} to the \gls{soa}.
        }%
        \label{tab:res:soac}
        \renewcommand*{\arraystretch}{2.3}
        \begin{threeparttable}
            \begin{tabular}{@{}ccccccccc@{}}
                \toprule
                \hol{\dmae} &
                \hol{Application} &
                \hol{Technology} &
                \htl{Supported}{Protocols} &
                \htl{Transfer}{Type} &
                \htl{Programming}{Model} &
                \hhl{Stream}{Modification}{Capability} &
                \htl{Modularity}{Configurability} &
                \hol{Area} \\
                \midrule
                \did{{CubeDMA}~\cite{fjeldtvedt_cubedma_2019}}{Fjeldtvedt~\etal} &
                \makecell[cc]{\revrep{h}{H}yperspectral \\ \revrep{i}{I}maging} &
                FPGA &
                \makecell[cc]{AXI~\tnote{a} \\ AXI4 Stream~\tnote{b}} &
                3D &
                \revrep{r}{R}egister \revrep{f}{F}ile &
                \revrep{n}{N}one &
                \makecell[cc]{\revrep{l}{L}imited \\ \revrep{c}{C}onf.} &
                \makecell[cc]{2162~LUTs \\ 1796~FFs} \\
                \did{{RDMA}~\cite{paraskevas_virtualized_2018}}{Paraskevas~\etal} &
                HPC &
                FPGA &
                AXI4 &
                \revrep{l}{L}inear &
                \makecell[cc]{\revrep{t}{T}ransfer \\ \revrep{d}{D}escriptors} &
                \revrep{n}{N}one &
                \revrep{n}{N}one &
                \revrep{n.a.}{N.A.} \\
                \did{{cDMA}~\cite{rhu_compressing_2018}}{Rhu~\etal} &
                DNN &
                ASIC~\tnote{c} &
                \revrep{n.a.}{N.A.} &
                \revrep{l}{L}inear &
                \revrep{n}{N}one &
                \makecell[cc]{\revrep{c}{C}ompression \& \\ \revrep{d}{D}ecompression} &
                \revrep{n}{N}o &
                \makecell[cc]{\SI{0.21}{\milli\metre\squared}~\tnote{d} \\ \SI{420}{\kilo\gateeq}~\tnote{e}} \\
                \dis{{Rossi~\etal~\cite{rossi2014ultra}}} &
                ULP &
                \makecell[cc]{\revrep{t}{T}ech.- \\ \revrep{i}{I}ndep.} &
                \makecell[cc]{STBus~\tnote{f} \\ OBI~\tnote{f}~\tnote{g}} &
                \revrep{l}{L}inear &
                \makecell[cc]{\revrep{per-pe}{Per-PE} \\ \revrep{r}{R}egister \revrep{f}{F}ile} &
                \revrep{n}{N}one &
                \makecell[cc]{\revrep{l}{L}imited \\ \revrep{c}{C}onfigurability} &
                \makecell[cc]{$\approx$ \SI{0.04}{\milli\metre\squared} \\ $\approx$ \SI{82}{\kilo\gateeq}} \\
                \did{{MT-DMA}~\cite{ma_mt-dma_2019}}{Ma~\etal} &
                \makecell[cc]{\revrep{s}{S}cientific \\ \revrep{c}{C}omputing} &
                ASIC &
                \revrep{c}{C}ustom &
                \makecell[cc]{2D \\ \revrep{a}{A}rb. \revrep{s}{S}trides} &
                \makecell[cc]{\revrep{t}{T}ransfer \\ \revrep{d}{D}escriptors} &
                \makecell[cc]{\revrep{b}{B}lock \\ \revrep{t}{T}ransp.} &
                \revrep{n}{N}one &
                \makecell[cc]{\SI{1.07}{\milli\metre\squared} \\ \SI{1.5}{\mega\gateeq}~\tnote{h}} \\
                \did{{FastVDMA}~\cite{noauthor_antmicro_2019}}{Antmicro} &
                \makecell[cc]{\revrep{g}{G}eneral- \\ purpose} &
                \makecell[cc]{\revrep{t}{T}ech.- \\ \revrep{i}{I}ndep.} &
                \makecell[cc]{AXI4~\tnote{f}~\tnote{i} \\ AXI4-Stream~\tnote{f}~\tnote{i} \\ Wishbone~\tnote{f}~\tnote{i}} &
                \revrep{l}{L}inear &
                \revrep{r}{R}egister \revrep{f}{F}ile &
                \revrep{n}{N}one &
                \makecell[cc]{\revrep{p}{P}rotocol \\ \revrep{s}{S}electable} &
                455~\revrep{s}{S}lices~\tnote{j} \\
                \did{{DMA-330}~\cite{arm_corelink_330_2012}}{ARM} &
                \makecell[cc]{\revrep{g}{G}eneral- \\ purpose} &
                \makecell[cc]{\revrep{t}{T}ech.- \\ \revrep{i}{I}ndep.} &
                \makecell[cc]{AXI3~\tnote{k} \\ \revrep{p}{P}eripheral \revrep{i}{I}ntf.~\tnote{l}} &
                \makecell[cc]{2D \\ \revrep{s}{S}catter-\revrep{g}{G}ather } &
                \makecell[cc]{\revrep{c}{C}ustom \\ \revrep{i}{I}nstructions } &
                \revrep{n}{N}one &
                \makecell[cc]{\revrep{y}{Y}es, \revrep{b}{B}ut \\ \revrep{n}{N}on-modular} &
                \revrep{n.a.}{N.A.} \\
                \did{{AXI DMA v7.1}~\cite{amd_xilinx_axi_2022}}{Xilinx} &
                \makecell[cc]{\revrep{g}{G}eneral- \\ purpose} &
                \makecell[cc]{FPGA \\ Xilinx-only} &
                \makecell[cc]{AXI4~\tnote{f} \\ AXI4-Stream~\tnote{f}} &
                \makecell[cc]{optional 2D \\ \revrep{s}{S}catter-\revrep{g}{G}ather } &
                \makecell[cc]{\revrep{t}{T}ransfer \\ \revrep{d}{D}escriptors} &
                \revrep{n}{N}one &
                \makecell[cc]{\revrep{y}{Y}es, \revrep{b}{B}ut \\ \revrep{n}{N}on-modular} &
                \makecell[cc]{2745~LUTs~\tnote{m} \\ 4738~FFs~\tnote{m} \\ \SI{216}{\kilo\bit}~BRAM~\tnote{m}} \\
                \did{{vDMA AXI}~\cite{plda_rambus_dma_nodate}}{RAMBUS} &
                \makecell[cc]{\revrep{g}{G}eneral- \\ purpose} &
                \makecell[cc]{\revrep{t}{T}ech.- \\ \revrep{i}{I}ndep.} &
                AXI3/4 &
                \makecell[cc]{2D \\ \revrep{s}{S}catter-\revrep{g}{G}ather } &
                \makecell[cc]{\revrep{t}{T}ransfer \\ \revrep{d}{D}escriptors} &
                \revrep{n}{N}one &
                \makecell[cc]{\revrep{y}{Y}es, \revrep{b}{B}ut \\ \revrep{n}{N}on-modular} &
                \revrep{n.a.}{N.A.} \\
                \did{{DW\_axi\_dmac}~\cite{synopsys_designware_nodate}}{Synopsys} &
                \makecell[cc]{\revrep{g}{G}eneral- \\ purpose} &
                \makecell[cc]{\revrep{t}{T}ech.- \\ \revrep{i}{I}ndep.} &
                \makecell[cc]{AXI3/4 \\ \revrep{p}{P}eripheral \revrep{i}{I}ntf.~\tnote{l}} &
                \makecell[cc]{2D \\ \revrep{s}{S}catter-\revrep{g}{G}ather } &
                \makecell[cc]{\revrep{t}{T}ransfer \\ \revrep{d}{D}escriptors} &
                \revrep{n}{N}one &
                \makecell[cc]{\revrep{y}{Y}es, \revrep{b}{B}ut \\ \revrep{n}{N}on-modular} &
                \revrep{n.a.}{N.A.} \\
                \did{{Dmaengine}~\cite{st_dmaengine_nodate}}{STMicroelectronics} &
                MCU &
                STM32 &
                \makecell[cc]{STBus~\tnote{n} \\ \revrep{p}{P}eripheral \revrep{i}{I}ntf.~\tnote{n}} &
                \revrep{l}{L}inear &
                \makecell[cc]{\revrep{c}{C}ustom \\ \revrep{i}{I}nstructions } &
                \revrep{n}{N}one &
                \revrep{n}{N}o &
                \revrep{n.a.}{N.A.} \\
                \did{{DDMA}~\cite{dcd_semi_nodate}}{DCD-SEMI} &
                MCU &
                \makecell[cc]{\revrep{t}{T}ech.- \\ \revrep{i}{I}ndep.~\tnote{n}} &
                \revrep{c}{C}ustom 32-bit~\tnote{o} &
                \revrep{l}{L}inear, fixed &
                \revrep{r}{R}egister \revrep{f}{F}ile &
                \revrep{n}{N}one &
                \revrep{n}{N}o &
                \revrep{n.a.}{N.A.} \\
                \did{{{\textmu}DMA}~\cite{pullini_dma_2017}}{Pullini~\etal} &
                MCU &
                \makecell[cc]{\revrep{t}{T}ech.- \\ \revrep{i}{I}ndep.} &
                \makecell[cc]{OBI~\tnote{f}~\tnote{g} \\ RX/TX \revrep{c}{C}hannels~\tnote{f}~\tnote{p}} &
                \revrep{l}{L}inear &
                \revrep{r}{R}egister \revrep{f}{F}ile &
                \revrep{n}{N}one &
                \makecell[cc]{\revrep{y}{Y}es, \revrep{b}{B}ut \\ \revrep{n}{N}on-modular} &
                \SI{15.4}{\kilo\gateeq} \\
                \dis{Morales~\etal~\cite{morales_low-area_2019}} &
                IoT &
                \makecell[cc]{\revrep{t}{T}ech.- \\ \revrep{i}{I}ndep.} &
                \makecell[cc]{AHB~\tnote{f} \\ \revrep{p}{P}erif. \revrep{i}{I}ntf.~\tnote{f}} &
                \revrep{l}{L}inear &
                \revrep{r}{R}egister \revrep{f}{F}ile &
                \revrep{n}{N}one &
                \makecell[cc]{\revrep{y}{Y}es, \revrep{b}{B}ut \\ \revrep{n}{N}on-modular} &
                \SI{3.2}{\kilo\gateeq} \\
                \dis{Su~\etal~\cite{su_processor-dma-based_2011}} &
                \makecell[cc]{\revrep{g}{G}eneral- \\ purpose} &
                \makecell[cc]{\revrep{t}{T}ech.- \\ \revrep{i}{I}ndep.~\tnote{n}} &
                AXI4 &
                \revrep{l}{L}inear~\tnote{n} &
                \revrep{r}{R}egister \revrep{f}{F}ile &
                \revrep{n}{N}one &
                \revrep{n}{N}o &
                \revrep{n.a.}{N.A.} \\
                \did{{VDMA}~\cite{nandan_high_2014}}{Nandan~\etal} &
                Video &
                \revrep{c}{C}ustom &
                \revrep{n.a.}{N.A.} &
                \makecell[cc]{2D \\ \revrep{a}{A}rb. \revrep{s}{S}trides} &
                \revrep{i}{I}ntegrated &
                \revrep{n}{N}one &
                \revrep{n}{N}o &
                \revrep{n.a.}{N.A.} \\
                \dis{Comisky~\etal~\cite{comisky_scalable_2000}} &
                MCU &
                \revrep{n.a.}{N.A.} &
                TR Bus &
                \revrep{l}{L}inear~\tnote{n} &
                \revrep{r}{R}egister \revrep{f}{F}ile &
                \revrep{n}{N}one &
                \revrep{n}{N}o &
                \revrep{n.a.}{N.A.} \\
                \midrule
                \dtw{This Work}{Architecture} &
                \makecell[cc]{\revrep{e}{E}xtreme-edge ULP, \\ \revrep{d}{D}atacenter HPC, \\ \revrep{a}{A}pplication-grade} &
                \makecell[cc]{\revrep{t}{T}ech.- \\ \revrep{i}{I}ndep.} &
                \makecell[cc]{AXI4, AXI4 Lite, \\ AXI4-Stream, \\ TL-UL, TL-UH, \\ OBI \\ Wishbone} &
                \makecell[cc]{\revrep{o}{O}ptional \revrep{N-D}ND \\ \revrep{a}{A}rb. \revrep{s}{S}trides \\ \revrep{s}{S}catter-\revrep{g}{G}ather} &
                \makecell[cc]{\revrep{r}{R}egister \revrep{f}{F}ile, \\ \revrep{t}{T}ransfer \revrep{d}{D}escriptors, \\ RISC-V ISA \revrep{e}{E}xt., \\ \revrep{c}{C}ustom} &
                \makecell[cc]{\revrep{m}{M}emory \revrep{i}{I}nit., \\ \revrep{i}{I}n-stream \\ \revrep{a}{A}ccelerator} &
                \makecell[cc]{\revrep{c}{C}onfigurable \\ and \revrep{m}{M}odular} &
                $\geq$ \SI{2}{\kilo\gateeq} \\
                \arrayrulecolor{black!30}\midrule
                \dtw{This Work}{{\occamy}} &
                \makecell[cc]{HPC \\ FP-\revrep{w}{W}orkloads} &
                ASIC~\tnote{q} &
                \makecell[cc]{AXI4 \\ OBI} &
                \makecell[cc]{2D \\ \revrep{a}{A}rb. \revrep{s}{S}trides \\ \revrep{s}{S}catter-\revrep{g}{G}ather} &
                RISC-V ISA ext. &
                \revrep{m}{M}emory \revrep{i}{I}nit. &
                \makecell[cc]{\revrep{c}{C}onfigurable \\ and \revrep{m}{M}odular} &
                $\approx$ \SI{75}{\kilo\gateeq} \\
                \dtw{This Work}{{\mempool}} &
                \makecell[cc]{\revrep{i}{I}mage \\ \revrep{p}{P}rocessing} &
                ASIC~\tnote{q} &
                \makecell[cc]{AXI4 \\ OBI~\tnote{r}} &
                \revrep{l}{L}inear &
                \revrep{r}{R}egister \revrep{f}{F}ile &
                \revrep{n}{N}one &
                \makecell[cc]{\revrep{c}{C}onfigurable \\ and \revrep{m}{M}odular} &
                $\approx$ \SI{45}{\kilo\gateeq} \\
                \dtw{This Work}{\pulpopen} &
                \makecell[cc]{ULP \\ ML} &
                \makecell[cc]{ASIC~\tnote{q} \\ FPGA~\tnote{q}} &
                \makecell[cc]{AXI4 \\ OBI} &
                \makecell[cc]{3D \\ \revrep{a}{A}rb. \revrep{s}{S}trides \\ \revrep{s}{S}catter-\revrep{g}{G}ather} &
                \revrep{r}{R}egister \revrep{f}{F}ile &
                \makecell[cc]{\revrep{b}{B}lock \\ \revrep{t}{T}ransp.} &
                \makecell[cc]{\revrep{c}{C}onfigurable \\ and \revrep{m}{M}odular} &
                $\approx$ \SI{50}{\kilo\gateeq} \\
                \dtw{This Work}{{\cheshire}} &
                \makecell[cc]{\revrep{a}{A}pplication- \\ \revrep{g}{G}rade} &
                ASIC~\tnote{q} &
                AXI4 &
                \revrep{l}{L}inear &
                \makecell[cc]{\revrep{t}{T}ransfer \\ \revrep{d}{D}escriptors} &
                \revrep{n}{N}one &
                \makecell[cc]{\revrep{c}{C}onfigurable \\ and \revrep{m}{M}odular} &
                $\approx$ \SI{60}{\kilo\gateeq} \\
                \dtw{This Work}{\controlpulp} &
                \makecell[cc]{Power \\ Management \\ MCU} &
                \makecell[cc]{\revrep{t}{T}ech.- \\ \revrep{i}{I}ndep.} &
                \makecell[cc]{AXI4 \\ OBI} &
                \makecell[cc]{3D \\ \revrep{a}{A}rb. \revrep{s}{S}trides \\ \revrep{s}{S}catter-\revrep{g}{G}ather} &
                \revrep{r}{R}egister \revrep{f}{F}ile &
                \revrep{n}{N}one &
                \makecell[cc]{\revrep{c}{C}onfigurable \\ and \revrep{m}{M}odular} &
                $\approx$ \SI{61}{\kilo\gateeq} \\
                \dtw{This Work}{IO-DMA} &
                \makecell[cc]{ULP \\ MCU} &
                \makecell[cc]{\revrep{t}{T}ech.- \\ \revrep{i}{I}ndep.} &
                \makecell[cc]{OBI} &
                \revrep{l}{L}inear &
                \revrep{r}{R}egister \revrep{f}{F}ile &
                \revrep{n}{N}one &
                \makecell[cc]{\revrep{c}{C}onfigurable \\ and \revrep{m}{M}odular} &
                $\approx$ \SI{2}{\kilo\gateeq} \\
                \arrayrulecolor{black!30}\bottomrule
            \end{tabular}
            \begin{tablenotes}[para, flushleft]
                \item[a] read-only
                \item[b] write-only
                \item[c] FreePDK45
                \item[d] \SI{28}{\nano\metre} node
                \item[e] assuming \SI{0.5}{\micro\metre\squared} per \SI{1}{\gateeq}
                \item[f] cross-protocol operation only
                \item[g] pre-1.0 version
                \item[h] assuming \SI{0.7}{\micro\metre\squared} per \SI{1}{\gateeq}
                \item[i] \emph{one read-only} and \emph{one write-only} protocol selectable
                \item[j] \SI{32}{\bit}, AXI4 read, AXI4-Stream write
                \item[k] one manager port, main interface
                \item[l] optional
                \item[m] \emph{UltraScale\_mm2s\_64DW\_1\_100 (xcku040, ffva1156, 1)}
                \item[m] to the best of our knowledge
                \item[o] wrapper for APB, AHB, AXI Lite available
                \item[p] very similar to OBI
                \item[q] main target
                \item[r] latency-tolerant version
            \end{tablenotes}
        \end{threeparttable}
    }
\end{table*}

We compare {\idma} to an extensive selection of commercial {\dma} solutions and {\dmaes} used in research platforms; an overview is shown in \cref{tab:res:soac}. %

In contrast to this work, existing {\dmaes} are designed for a given system, a family of systems, or even a specific application on a system. %
These engines lack modularity and cannot be readily retargeted to a different system. %
To the best of our knowledge, our work is the first fully modular and universal {\dmaa}. %
Moreover, most of the \revrep{DMAs}{\dmaes} in our comparison are closed-source designs and thus not accessible to the research community, hindering or even preventing  benchmarking and quantitative comparisons. %

We identify two general categories of {\dmaes}: large high-bandwidth engines specialized in efficient memory transfers and low-footprint engines designed for accessing peripherals efficiently. %
Ma~\etal~\cite{ma_mt-dma_2019}, Paraskevas~\etal~\cite{paraskevas_virtualized_2018}, and Rossi~\etal~\cite{rossi2014ultra} present high-performance \revrep{DMA units}{\dmaes} ranging in size from \SI{82}{\kilo\gateeq} to over \SI{1.5}{\mega\gateeq}. %
On the contrary, {\dmaes} designed for accelerating accesses to chip peripherals, as shown in the works of Pullini~\etal~\cite{pullini_dma_2017} and Morales~\etal~\cite{morales_low-area_2019} trade-off performance for area efficiency by minimizing buffer space~\cite{pullini_dma_2017} and supporting only simpler on-chip protocols like AHB~\cite{morales_low-area_2019} or OBI~\cite{pullini_dma_2017}. %
Our \revrep{DMA architecture}{\idma} can be parameterized to achieve peak performance as a high-bandwidth engine in \gls{hpc} systems, see \Cref{sec:sys:occamy}, as well as to require less area ($<$\SI{2}{\kilo\gateeq}) than the ultra-lightweight design of Pullini~\etal's \emph{{\textmu}DMA}~\cite{pullini_dma_2017}. %

{\dmaes} can be grouped according to their system binding: register, transfer-descriptor, and instruction-based. %
Engines requiring a high degree of agility~\cite{fjeldtvedt_cubedma_2019, rossi2014ultra} or featuring a small footprint~\cite{morales_low-area_2019, pullini_dma_2017, noauthor_antmicro_2019} tend to use a register-based interface. %
\Glspl{pe} write the transfer information in a dedicated register space and use a read or write operation to a special register location to launch the transfer. %
In more memory-compute-decoupled systems~\cite{paraskevas_virtualized_2018} or manycore environments~\cite{paraskevas_virtualized_2018, ma_mt-dma_2019} transfer descriptors prevail. %
In some \gls{mcu} platforms~\cite{st_dmaengine_nodate, arm_corelink_330_2012} \revrep{DMAs}{\dmaes} are programmed using a custom instruction stream. %
Generally, \revrep{DMA units}{\dmaes} only feature one programming interface with some exceptions: both Xilinx's \emph{AXI DMA v7.1} and Synopsys' \emph{DW\_axi\_dmac} support next to their primary transfer-descriptors-based also a register-based interface usable with only a reduced subset of the engines' features~\cite{amd_xilinx_axi_2022, synopsys_designware_nodate}. %
Our flexible architecture allows us to use these three system bindings without limiting {\idma}'s capabilities. %
With our standardized interfaces, any custom binding can be implemented, fully tailoring the engine to the system it is attached to. %
Compared to the prevailing approach of using custom \glspl{isa}~\cite{st_dmaengine_nodate, arm_corelink_330_2012}, our \emph{inst\tus64} {\fe} extends the RISC-V \gls{isa}, allowing extremely agile programming of complex transfer patterns. %

\gls{soc} {\dmaes}, e.g., Fjeldtvedt~\etal~\cite{fjeldtvedt_cubedma_2019}, Rossi~\etal~\cite{rossi2014ultra}, and Morales~\etal~\cite{morales_low-area_2019} feature a fixed configuration of on-chip protocol(s). %
Some controllers allow selectively adding a simple interface to connect to peripherals. %
The {\dma} \glspl{ip} from Synopsys~\cite{synopsys_designware_nodate} and ARM~\cite{arm_corelink_330_2012} are two examples. %
We identify one exception to this rule\revrep{,}{:} \emph{FastVDMA} from {Antmicro}~\cite{noauthor_antmicro_2019} can be configured to select one read- and one write-only port from a selection of three protocols. %
\emph{FastVDMA} only supports unidirectional data flow from one read to the other write port. %
\emph{Inter-port} operation, meaning copy\rev{ing} data from one port and storing it using the same port, is not supported, which can limit its usability. %
{\idma} allows the selective addition of one or multiple read or write interface ports from a list of currently five industry-standard on-chip protocols. %
If configured, our engines allow bidirectional data movement in \emph{inter-port} and \emph{intra-port} operations. %
Thanks to the standardization of interfaces and the separation of data movement from protocol handling, new on-chip protocols can be added quickly by implementing \emph{at most} three modules, each only a couple of hundred \si{\gateeq}s of complexity. %

Many {\dmaes} support transfers with more than one addressing dimension. %
Two-dimensional transfers are commonly accelerated in hardware~\cite{ma_mt-dma_2019,  arm_corelink_330_2012, amd_xilinx_axi_2022}. %
Fjeldtvedt~{\etal}'s \emph{CubeDMA} can even handle three-dimensional transfers. %
Any higher dimensional transfer is handled in software either by repetitively launching simpler transfers~\cite{fjeldtvedt_cubedma_2019}~\cite{rossi2014ultra} or by employing transfer descriptor chaining~\cite{ma_mt-dma_2019, amd_xilinx_axi_2022}. %
Our \emph{tensor{\tus}ND} {\me} can execute arbitrary high dimensional transfers in hardware. %
Our \emph{desc\tus64} {\fe} supports descriptor chaining to handle arbitrarily-shaped transfers without putting any load on the \gls{pe}. %
Our flexible architecture can easily accelerate special transfer patterns required by a specific application: once programmed, our novel \emph{rt{\tus}3D} {\me} autonomously fetches strided sensor data without involving any \gls{pe}. %
Rhu~\etal~\cite{rhu_compressing_2018} and Ma~\etal~\cite{ma_mt-dma_2019} present {\dmaes} able to modify the data while it is copied. %
Although their work proposes a solution for their respective application space, none of these engines present a standardized interface to exchange \emph{stream acceleration modules} between platforms easily. %
Additionally, both engines, especially \emph{MT-DMA}, impose substantial area overhead, limiting their applicability in \gls{ulp} designs. %
Our engines feature a well-defined interface accessing the byte stream while data is copied, allowing us to include existing accelerators. %
Furthermore, we provide a novel, ultra-lightweight memory initialization feature, typically requiring less than \SI{100}{\gateeq}. %

\emph{FastVDMA}~\cite{noauthor_antmicro_2019} shows basic modularity by allowing \revrep{the user}{users} to select one read and one write protocol from a list of three on-chip protocols. %
Its modularity is thus limited to the {\be}, and there is neither any facility to change between different programming interfaces nor a way of easily adding more complex multi-dimensional affine stream movement support. %
As presented in this work, our approach tackles these limitations by specifying and implementing the first fully modular, parametric, universal {\dmaa}. %

\section{Conclusion}
\label{sec:conclusion}

We present {\idma}, a modular, highly parametric {\dmaa} composed of three parts ({\fe}, {\me}, and {\be}), allowing our engines to be customized to suit a wide range of systems, platforms, and applications. %
We showcase its adaptability and real-life benefits by integrating it into five systems targeting various applications. %
In a minimal Linux-capable \gls{soc}, our architecture increases bus utilization by up to 6\x~while reducing \gls{dmae} resource footprint by more than \SI{10}{\percent}. %
In a \gls{ulp} edge-node system, we improve MobileNetV1 inference performance from  \SI{7.9}{MAC\per cycle} to \SI{8.3}{MAC\per cycle} while reducing the compute cluster area by \SI{10}{\percent} compared to the baseline {MCHAN}~\cite{rossi2014ultra}. %
We demonstrate {\idma}'s applicability in real-time systems by introducing a small \emph{real-time \me} requiring only \SI{11}{\kilo\gateeq} while completely liberating the core from any periodic sensor polling. %

We demonstrate speedups of up to 8.4\x~and 15.8\x~in high-performance manycore systems designed for floating-point and integer workloads, respectively, compared to their baselines without {\dmaes}. %
We evaluate the area, timing, latency, and performance of {\idma}, resulting in area and timing models that allow us to estimate the synthesized area and timing characteristics of any parameterization within \SI{9}{\percent} of the actual result. %

Our architecture enables the creation of both ultra-small {\idmaes} incurring less than \SI{2}{\kilo\gateeq}, as well as large high-performance {\idmaes} running at over \SI{1}{\giga\hertz} on a \SI{12}{\nano\metre} node. %
Our {\bes} incur only two cycles of latency from accepting an \revrep{N-dimensional}{ND} transfer descriptor to the first read request being issued on the engine's protocol port. %
They show high agility, even in ultra-deep memory systems. %
Flexibility and parameterization allow us to create configurations that achieve asymptotically full bus utilization and can fully hide latency in arbitrary deep memory systems while incurring less than \SI{400}{\gateeq} per trackable outstanding transfer. %
In a \SI{32}{\bit} system, our {\idmaes} achieve almost perfect bus utilization for \SI{16}{\byte}-long transfers when accessing an endpoint with 100 cycles of latency. %
The synthesizable \gls{rtl} description of {\idma} is available free and open-source. %

\section*{Acknowledgments}
We thank %
Fabian Schuiki,
Florian Zaruba,
Kevin Schaerer, %
Axel Vanoni, %
Tobias Senti, and %
Michele Raeber %
for their valuable contributions to the research project. %
This work was supported in part through funding from the %
European High Performance Computing Joint Undertaking (JU) %
under Framework Partnership Agreement No 800928 and %
Specific Grant Agreement No: 101036168 (EPI SGA2) and No: 101034126 (The EU Pilot). %
\revdel{%
The JU receives support from the European Union's Horizon 2020 research and innovation program and %
Spain, Italy, Switzerland, Netherlands, Portugal, Germany, France, Greece, Sweden, Croatia, and Turkey. %
}
%

%


%
\newcommand{\lucaphd}{He is currently pursuing a Ph.D. degree in the Digital Circuits and Systems group of Prof.\ Benini.}
\newcommand{\ethgrad}[2]{received his BSc and MSc degrees in electrical engineering and information technology from ETH Zurich in #1 and #2, respectively.}
\newcommand{\researchinterests}[1]{His research interests include #1.}

\begin{IEEEbiography}[%
    {\includegraphics[width=1in,height=1.25in,clip,keepaspectratio]{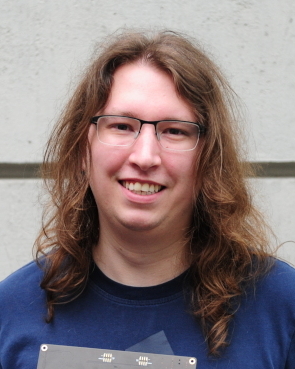}}%
    ]{Thomas Benz}
    \ethgrad{2018}{2020}
    \lucaphd{}
    \researchinterests{energy-efficient high-performance computer architectures and the design of \acrshortpl{asic}}
\end{IEEEbiography}

\begin{IEEEbiography}[%
    {\includegraphics[width=1in,height=1.25in,clip,keepaspectratio]{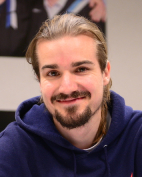}}%
    ]{Michael Rogenmoser}
    \ethgrad{2020}{2021}
    \lucaphd{}
    \researchinterests{%
        fault-tolerant processing architectures and multicore heterogeneous \acrshortpl{soc} for space}
\end{IEEEbiography}

\begin{IEEEbiography}[%
    {\includegraphics[width=1in,height=1.25in,clip,keepaspectratio]{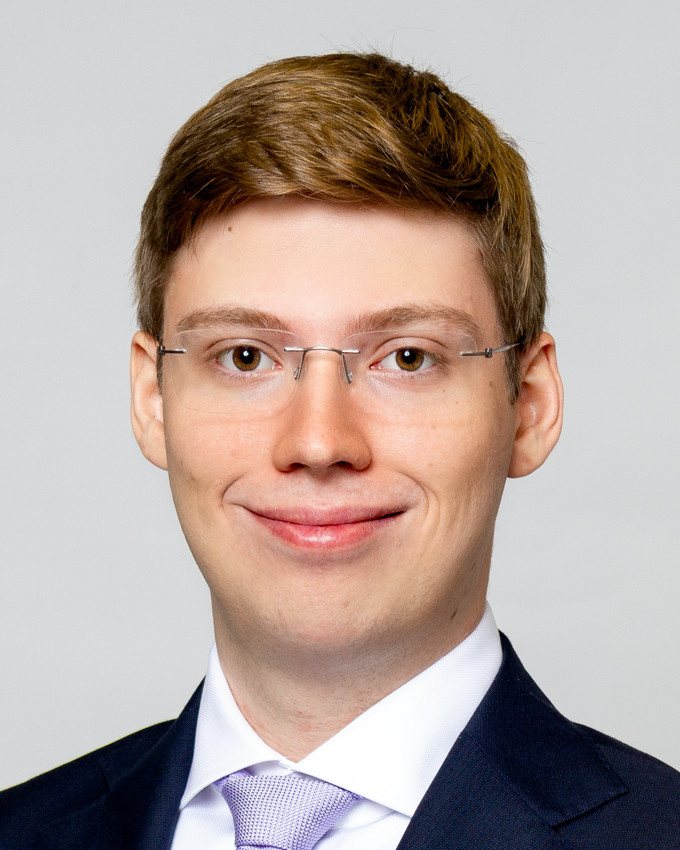}}%
    ]{Paul Scheffler}
    \ethgrad{2018}{2020}
    \lucaphd{}
    \researchinterests{%
        acceleration of sparse and irregular workloads, on-chip interconnects, manycore architectures, and high-performance computing}
\end{IEEEbiography}

\ifx\showrebuttal\undefined
    \vfill
    \newpage
    \vfill
\fi

\begin{IEEEbiography}[%
    {\includegraphics[width=1in,height=1.25in,clip,keepaspectratio]{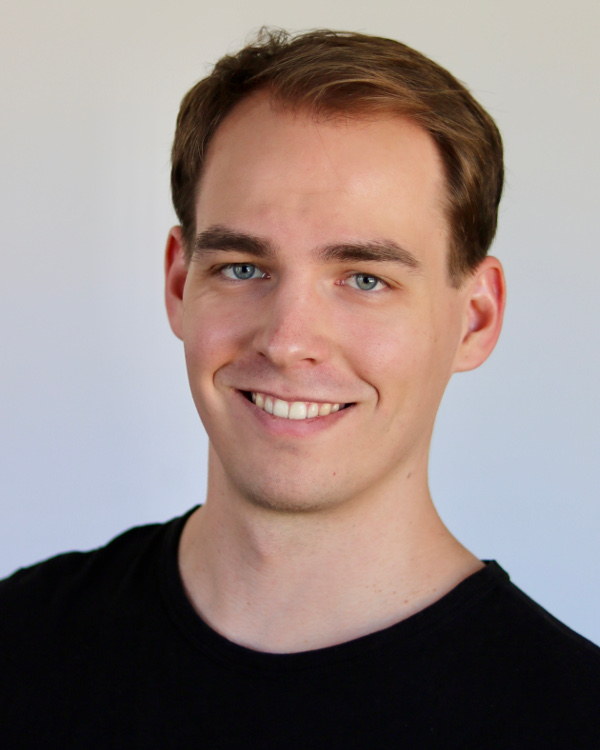}}%
    ]{Samuel Riedel}
    \ethgrad{2017}{2019}
    \lucaphd{}
    \researchinterests{%
        computer architecture, focusing on manycore systems and their programming model}
\end{IEEEbiography}

\begin{IEEEbiography}[%
    {\includegraphics[width=1in,height=1.25in,clip,keepaspectratio]{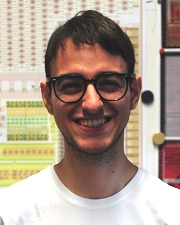}}%
    ]{Alessandro Ottaviano}
    received the B.Sc. in Physical Engineering from Politecnico di Torino, Italy, and the M.Sc. in Electrical Engineering as a joint degree between Politecnico di Torino, Grenoble INP-Phelma and EPFL Lausanne, in 2018 and 2020 respectively. %
    \lucaphd{}
    \researchinterests{%
      real-time and predictable computing systems and energy-efficient processor
      architecture}
\end{IEEEbiography}

\begin{IEEEbiography}[%
    {\includegraphics[width=1in,height=1.25in,clip,keepaspectratio]{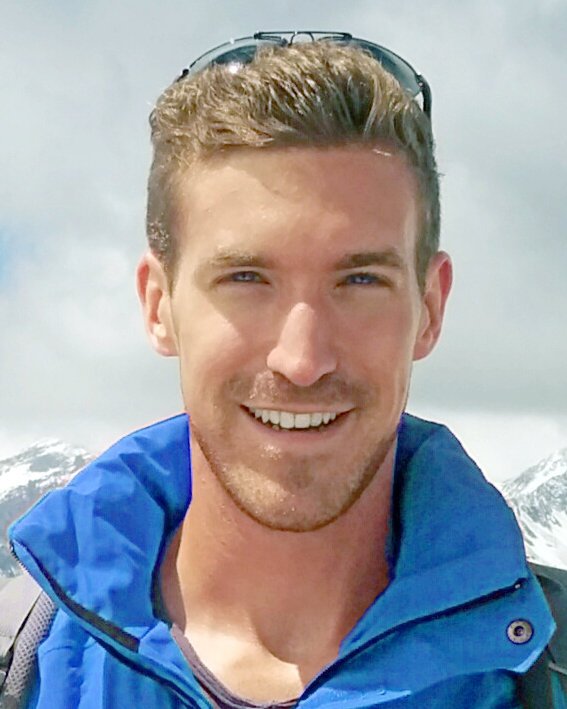}}%
    ]{Andreas Kurth}
    \ethgrad{2014}{2017}
    He completed his Ph.D. in the Digital Circuits and Systems group of Prof.\ Benini in 2022. %
    \researchinterests{%
        the architecture and programming of heterogeneous \acrshortpl{soc} and %
        accelerator-rich computing systems}
\end{IEEEbiography}

\begin{IEEEbiography}[%
    {\includegraphics[width=1in,height=1.25in,clip,keepaspectratio]{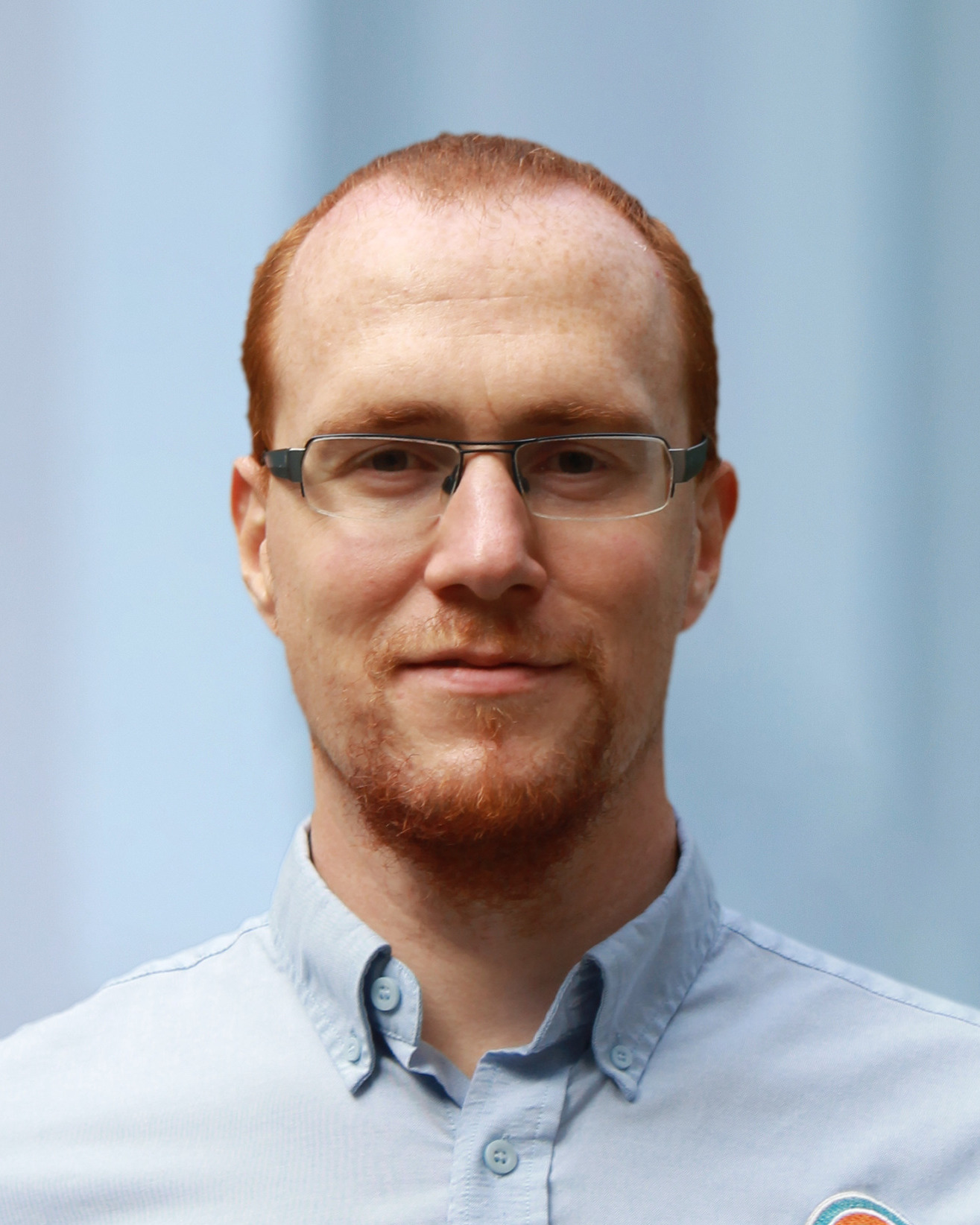}}%
    ]{Torsten Hoefler}
    is a Professor of Computer Science at ETH Zürich, Switzerland. %
    He received his PhD from Indiana University. %
    Dr. Hoefler's interests are around performance-centric software and hardware development. %
    He is a Fellow of the ACM and a member of the Academia Europaea.
\end{IEEEbiography}

\begin{IEEEbiography}[%
    {\includegraphics[width=1in,height=1.25in,clip,keepaspectratio]{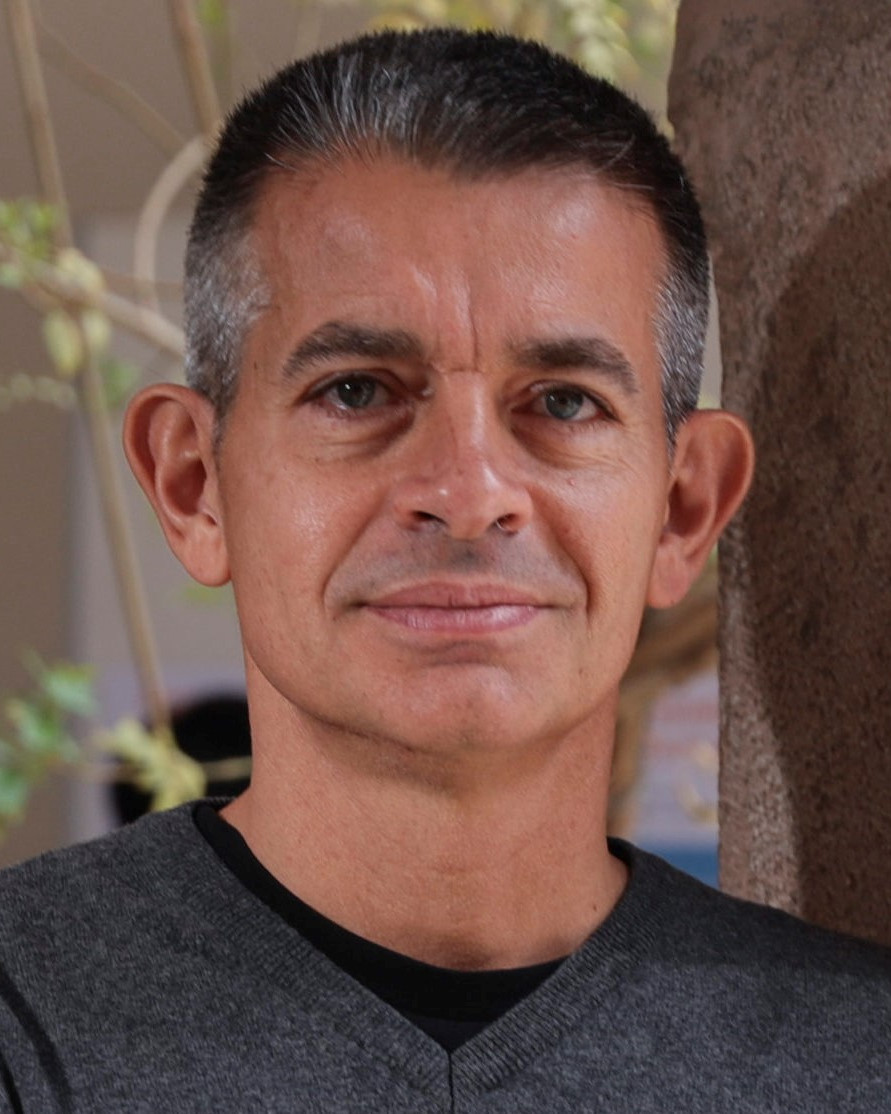}}%
    ]{Luca Benini}
    (F'07) holds the chair of Digital Circuits and Systems at ETH Zurich and is Full Professor at the Università di Bologna.
    Dr.\ Benini’s research interests are in energy-efficient computing systems design, from embedded to high-performance.
    He has published more than 1000 peer-reviewed papers and five books.
    He is a Fellow of the ACM and a member of Academia Europaea.
\end{IEEEbiography}

\vfill
\end{document}